\documentclass[preprint,aps,amsmath,amssymb,prd]{revtex4-1}
\usepackage{bm}
\usepackage{graphicx}
\usepackage{booktabs}
\usepackage{slashed}
\usepackage{hyperref}
\usepackage{amssymb,amsmath,amsfonts}
\usepackage{color}
\usepackage[utf8]{inputenc}
\usepackage[caption=false]{subfig}

\newcommand{\frapq}{\frac{p}{Q}}
\newcommand{\frasqpq}{\frac{p^2}{Q^2}}
\newcommand{\frasqqp}{\frac{Q^2}{p^2}}
\newcommand{\fraqupq}{\frac{p^4}{Q^4}}

\newcommand{\fraqp}{\frac{Q}{p}}
\newcommand{\be}{\begin{equation}}
\newcommand{\ee}{\end{equation}}
\newcommand{\bea}{\begin{eqnarray}}
\newcommand{\eea}{\end{eqnarray}}

\renewcommand{\vec}[1]{{\bm #1}}

\begin{document}

\title{Effective models of a semi-quark gluon plasma}

\author{Yoshimasa Hidaka}
\affiliation{KEK Theory Center, Tsukuba 305-0801, Japan}
\affiliation{Graduate University for Advanced Studies (Sokendai), Tsukuba 305-0801, Japan}
\affiliation{RIKEN iTHEMS, RIKEN, Wako 351-0198, Japan}

\author{Robert D. Pisarski}
\affiliation{Department of Physics, Brookhaven National Laboratory, Upton, NY 11973}

\begin{abstract}
  In the deconfined regime of a non-Abelian gauge theory at nonzero temperature, previously
  it was argued that if a (gauge invariant) source is added to generate nonzero holonomy,
  that this source must be linear for small holonomy.  The simplest example of
  this is the second Bernoulli polynomial.
  However, then there is a conundrum in computing the free energy to $\sim g^3$ in the coupling
  constant $g$, as part of the free energy is discontinuous as the holonomy vanishes.
  In this paper we investigate two ways of generating the second Bernoulli polynomial dynamically:
  as a mass derivative of an auxiliary field, and from 
  two dimensional ghosts embedded isotropically in four dimensions.   Computing
  the holonomous hard thermal loop (HHTL) in the gluon self-energy, we find that the limit of small holonomy
  is only well behaved for two dimensional ghosts, with a free energy which 
  to $\sim g^3$ is continuous as the holonomy vanishes.
\end{abstract}
\preprint{KEK-TH-2252, J-PARC-TH-224}

\maketitle
\newpage

\section{Introduction}
\label{sec:introduction}

The behavior of gauge theories is intrinsically of fundamental interest, and especially for understanding the behavior of
collisions of heavy ions at ultrarelativistic energies.
Coming down from high temperature,
resummations of perturbation theory can be used down to several times the transition temperature
\cite{Kajantie:2002wa,Haque:2012my,Haque:2013sja,Andersen:2015eoa};
coming up from low temperature, hadronic gas models are useful.  This leaves the most interesting region,
from the transition temperature to a few times that
\cite{Gross:1980br,Weiss:1980rj,Belyaev:1991gh,Belyaev:1991np,Smilga:1993vb,Kogan:1993bz,bhattacharya_interface_1991,bhattacharya_zn_1992,Sawayanagi:1992as,Sawayanagi:1994nz,Skalozub:1992un,Skalozub:1992pb,Skalozub:1994br,Skalozub:1994pa,Borisenko:1994jn,Skalozub:2004ab,Bordag:2018aii,Borisenko:2020dej,Skalozub:2020idf,KorthalsAltes:1993ca,Giovannangeli:2002uv,Giovannangeli:2004sg,dumitru_dense_2005,oswald_beta-functions_2006,Hidaka:2008dr,Hidaka:2009hs,Hidaka:2009xh,Hidaka:2009ma,Dumitru:2010mj,Dumitru:2012fw,Kashiwa:2012wa,Pisarski:2012bj,Dumitru:2013xna,smith_effective_2013,Lin:2013qu,Gale:2014dfa,Guo:2014zra,Hidaka:2015ima,Pisarski:2016ixt,Guo:2018scp,Pisarski:2012bj,Nishimura:2017crr,Folkestad:2018psc,KorthalsAltes:2019yih,KorthalsAltes:2020ryu,altes_nishimura,hidaka_new,Guo:2020jvc}.
This has been described as a semi-quark gluon plasma \cite{Hidaka:2008dr,Hidaka:2009hs,Hidaka:2009xh,Hidaka:2009ma},
when the deconfined phase exhibits nontrivial holonomy.

It is most direct to use an effective model, where the nontrivial holonomy is generated by adding a term
to the action by hand.  Given the wealth of results from numerical simulations on the lattice \cite{Philipsen:2019rjq},
it is relatively straightforward to construct models which well fit the pressure and related thermodynamic
quantities \cite{Dumitru:2010mj,Dumitru:2012fw}.

Previous studies have suggested that in constructing effective theories, that even at the classical level,
one has to ensure that there is not a transition between the strict perturbative regime, where the holonomy vanishes,
and that with nonzero holonomy.  While there is no strict order parameter between these two
regimes, there can be a first order transition.  For the theory in four spacetime dimensions at nonzero temperature,
such a first order transition, {\it in} the deconfined phase, appears generically
\cite{Dumitru:2012fw,KorthalsAltes:2020ryu}.

The reason for this is easy to understand.  
There is no potential for the holonomy classically, while a potential
is generated at one-loop order.
The eigenvalues of the thermal Wilson line are gauge invariant, and it is natural
to consider their logarithm, $q$.  
For a $SU(2)$ gauge group there is a single $q$, with more $q$'s for larger gauge groups.
To lowest order, $q$ is proportional to the static component of the gauge field $A_0$, although
the relation becomes more involved to higher order \cite{Belyaev:1991gh}.
As the logarithm of an exponential, the $q$'s are periodic variables, defined properly in the Weyl chamber
for the Lie algebra of the gauge group.

Ignoring these technicalities, to leading order the potential for the holonomy $q$ is elementary,
proportional to the fourth Bernoulli polynomial,
\begin{equation}
  {\cal V}_{\rm pert}(q) \sim + \, T^4 \, q^2 (1 - q)^2 \; ,
\end{equation}
where $T$ is the temperature, and $q$ is a periodic variable, here defined for $q: 0 \rightarrow 1$.
This form is unchanged to two-loop order \cite{Belyaev:1991gh,bhattacharya_interface_1991,bhattacharya_zn_1992,KorthalsAltes:1993ca,Giovannangeli:2002uv,Giovannangeli:2004sg,Dumitru:2013xna,Guo:2014zra,Guo:2018scp}.
The term $\sim +\, q^2 \sim {\rm tr} \, A_0^2$ just represents Debye screening of static electric fields.
The term $\sim q^4 \sim {\rm tr} \, A_0^4$ is also unexceptional, a type of induced Higgs coupling for static electric fields,
familiar from a perturbative analysis \cite{Kajantie:2002wa}.
What is striking is the cubic term: this is given by integrating over the mode with zero energy,
and  enters with a {\it negative} sign.
As is typical of mean field theory, such a cubic term automatically generates a first order transition.
This was first suggested in Ref. \cite{Dumitru:2012fw}, and analyzed in detail in Ref. \cite{KorthalsAltes:2020ryu}.

We stress that this first order transition would occur {\it in} the deconfined phase, at temperatures
{\it above} that for the deconfining and/or chiral symmetry transitions.
There is absolutely
no hint of any such transition from numerical simulations on the lattice \cite{Philipsen:2019rjq}.
The simplest way of avoiding such an unwanted transition is to add a term which is linear in $q$ for small $q$
\cite{Dumitru:2010mj,Dumitru:2012fw,KorthalsAltes:2020ryu}, so that the expectation value of $q$ is always nonzero.
The simplest choice is the second Bernoulli polynomial,
\begin{equation}
{\cal V}_{\rm non-pert}(q) =  C \, T^2 \, B_2(q) \sim - \, C \, T^2 \; q (1-q) \; ,
  \label{potential_b2}
\end{equation}
where $C$ has dimensions of mass squared.

In previous analysis, the coefficient $C$ was taken to be constant,
so that the nonperturbative term in Eq.~(\ref{potential_b2}) contributes to the pressure $\sim T^2$.
In the pure glue theory, that the leading power law correction to the pressure of an
ideal gas is proportional to $T^2$ has been found to be valid, to a good approximation, in both
$2+1$ \cite{Caselle:2011mn,Bicudo:2013yza,Bicudo:2014cra} and $3+1$ dimensions
\cite{Beinlich:1996xg,Pisarski:2000eq,Meisinger:2001cq,Pisarski:2006hz,Datta:2010sq,Borsanyi:2012ve,Lucini:2012gg,Lucini:2013qja,Kitazawa:2016dsl,Giusti:2016iqr,Kitazawa:2016dsl,Iritani:2018idk,Hirakida:2018uoy,Caselle:2018kap}.
Since it has dimensions of mass (squared),
this term is manifestly nonperturbative.  At high $T$, $q \sim C/T^2$, so that $q \neq 0$
at {\it any} temperature.  Consequently, the quark gluon plasma is {\it always} holonomous, even if
the holonomy is infinitesimally small at high temperature.
In a pure gauge theory, that $\langle q \rangle \neq 0$ when $C \neq 0$ has been demonstrated
carefully for both two and an infinite number of colors \cite{KorthalsAltes:2020ryu}, but because of the
cubic term in the perturbative potential, it almost certainly applies for any gauge group.

(With dynamical quarks, it is no longer true that the leading power law correction to the pressure is $\sim T^2$.
Nevertheless, an effective model, where dynamical quarks are folded into an effective theory with
a nonperturbative term like Eq.~(\ref{potential_b2}), gives a reasonable description of
the pressure \cite{Pisarski:2016ixt,Folkestad:2018psc} without the introduction of new parameters.  This is done by keeping
$T_c$ in the gluonic part of the effective potential the {\it same} as in the pure glue theory, with the
temperature for the chiral phase transition, which is $\approx T_c/2$, 
arises by adjusting a Yukawa coupling in the coupling between quarks
and effective meson degrees of freedom.)

A related problem is the behavior of the free energy in the presence of gauge invariant sources
\cite{KorthalsAltes:2019yih,KorthalsAltes:2020ryu}.
For any source which is a sum over a finite number of Polyakov loops, because of the
term $\sim - q^3 $ in ${\cal V}_{\rm pert}(q)$, there is a first order transition
between the phase with zero and nonzero holonomy.  A source proportional to the second Bernoulli polynomial avoids this.

Doing so, however, a conundrum arises.  A well defined and gauge invariant quantity is the free energy,
computed perturbatively.  The free energy $\sim 1$ and $\sim g^2$ is sensible, but
for weak holonomy, that $\sim g^3$ is discontinuous as the
holonomy vanishes.  This discontinuity
occurs for {\it any} gauge invariant source, and follows
directly from the equation of motion which the source must satisfy
\cite{KorthalsAltes:2019yih,KorthalsAltes:2020ryu}.
This discontinuity is unexpected, and most uncharacteristic of ordinary theories in the presence of external sources,
whose effects smoothly vanish as the source does.

We contrast this with the behavior of gauge theories when compactified in one spatial direction to a size
where the gauge coupling is small (``femto-torus'')
\cite{Unsal:2008ch,Poppitz:2008hr,Shifman:2009tp,Golkar:2009aq,Nishimura:2011md,Poppitz:2012sw,Poppitz:2012nz,Anber:2013doa,Poppitz:2013zqa,Ogilvie:2014bwa,Kanazawa:2017mgw}.
This often induces nonzero holonomy in the compactified direction, as semiclassical configurations such as
magnetic monopoles and bions arise dynamically.
For small spatial directions, however, there is no sign that the associated free energy exhibits
any discontinuity.  Unfortunately, it is not possible to study the theory analytically
as the size of the compactified dimension
becomes large.  

The object of the present study is to see if the conundrum in the presence of external sources
\cite{KorthalsAltes:2019yih,KorthalsAltes:2020ryu} can be avoided by generating $B_2(q)$ dynamically,
through the introduction of auxiliary fields.  The presence of these auxiliary fields can 
be viewed as a caricature of the non-perturbative physics which generates nontrivial holonomy
at nonzero temperature 
\cite{Unsal:2008ch,Poppitz:2008hr,Shifman:2009tp,Golkar:2009aq,Nishimura:2011md,Poppitz:2012sw,Poppitz:2012nz,Anber:2013doa,Poppitz:2013zqa,Ogilvie:2014bwa,Kanazawa:2017mgw}.
At high temperature, when the holonomy is infinitesimally small, it is then sensible to ask if the free energy is
smoothly behaved as $T \rightarrow \infty$.

There are at least two ways of generating the second Bernoulli polynomial by the introduction of 
an auxiliary field, which is assumed to lie in the adjoint representation.
The first is to introduce a mass for additional field,
and then take a derivative with respect to the mass, Sec. (\ref{sec:massive}); see, {\it e.g.},
Refs. \cite{Meisinger:2001cq,Poppitz:2012sw,Poppitz:2012nz}.  The second
way is to embed fields in two spacetime dimensions isotropically in four dimensions, Sec. (\ref{sec:two_dim}).
We consider hard thermal loops (HTLs) \cite{Bellac:2011kqa} at nonzero holonomy
\cite{Hidaka:2008dr,Hidaka:2009hs,Hidaka:2009xh,Hidaka:2009ma}, which we term
holonomous hard thermal loops (HHTL).  
In Sec. (\ref{sec:hhtl}) we compute the HHTL in the effective gluon propagator for both fields in the Euclidean theory.
At zero holonomy this is just the Debye mass squared, but in the static limit at nonzero holonomy,
the HHTLs are nontrivial functions of the dimensionless ratio between the holonomy and the spatial momentum,
both of which must be soft, $\sim g T$.
Surprisingly, the HHTL propagator
for the massive auxiliary field has terms which are not $\sim g^2$ for small holonomy, as one would expect, but $\sim g$.
This is not consistent with a smooth approach to the perturbative limit.  In contrast, the two dimensional fields
give a HHTL propagator for which the leading corrections are $\sim g^2$.

We then use this to compute the corrections to the free energy to $\sim g^3$ at nonzero holonomy in Sec. (\ref{sec:free_energy}),
and show that the contribution vanishes smoothly as the holonomy
vanishes.  This solves the conundrum when the holonomy is generated by external sources
\cite{KorthalsAltes:2019yih,KorthalsAltes:2020ryu}, and agrees with the results
on a femto-torus
\cite{Unsal:2008ch,Poppitz:2008hr,Shifman:2009tp,Golkar:2009aq,Nishimura:2011md,Poppitz:2012sw,Poppitz:2012nz,Anber:2013doa,Poppitz:2013zqa,Ogilvie:2014bwa,Kanazawa:2017mgw}.

Further, it is trivial to generalize the HHTL gluon propagators to Minkowski spacetime.  Thus
this effective theory allows the computation of quantities as transport coefficients using the effective models of
Refs. \cite{Hidaka:2008dr,Hidaka:2009hs,Hidaka:2009xh,Hidaka:2009ma,Dumitru:2010mj,Dumitru:2012fw,Kashiwa:2012wa,Pisarski:2012bj,Dumitru:2013xna,smith_effective_2013,Lin:2013qu,Gale:2014dfa,Guo:2014zra,Hidaka:2015ima,Pisarski:2016ixt,Guo:2018scp,Pisarski:2012bj,Nishimura:2017crr,hidaka_new},
which are being carried out \cite{hidaka_new}.

\section{Auxiliary massive fields}
\label{sec:massive}

We start with an auxiliary massive field in the adjoint representation, and take the derivative with respect to the mass squared:
\begin{equation}
\begin{split}
S_\text{m}= C \left.\frac{\partial}{\partial m^{2}}\mathrm{Tr} \ln (-D_{\mu}^{2}+m^{2})\right|_{m^{2}=0} \; ,
\end{split}
\label{mass_action}
\end{equation}
where $D_\mu=\partial_\mu-igA_\mu$ is the covariant derivative with the gluon field $A_\mu$.
$C$ has dimensions of mass squared, which may depend on temperature, and the traces are over both spacetime
and color.  We stress that taking a derivative with respect to mass is nothing more than 
a mathematical device to give us the desired result,
an effective potential proportional to the second Bernoulli polynomial.

\subsection{Effective potential at nonzero holonomy}

To compute at nonzero holonomy, we take a background gauge potential
\begin{equation}
  A_0^{{\rm cl}, a b} = \frac{Q^a}{g} \delta^{a b} \;\; , \;\; Q^a = 2 \pi T q^a \; ,
\end{equation}
where from $Q^a$ we pull out factors to introduce the dimensionless $q^a$.
The $Q^a$'s are diagonal elements of a
$SU(N)$ matrix, and so are traceless, $\sum_{a=1}^N Q^a = 0$.  We write the
adjoint representation as a two index tensor over fundamental indices, and so the projector
\begin{equation}
  \mathcal{P}^{ab}_{dc} = \delta^{a}_d \delta^b_c - \frac{1}{N} \; \delta^{ab}\delta_{cd} \; 
\end{equation}
often enters; $\mathcal{P}^{ab}_{dc} = \mathcal{P}^{ab,cd}$  \cite{Hidaka:2008dr,Hidaka:2009hs,Hidaka:2009xh,Hidaka:2009ma}.

For massless fields in two and four spacetime dimensions the first four Bernoulli polynomials arise naturally at one
loop order,
\begin{eqnarray}
 B_1(q) &=& -\frac{1}{2} + q \; , \nonumber\\
  B_2(q)&=& \frac{1}{6} - q(1-q) \; , \nonumber \\
  B_3(q)& = & \frac{1}{2} \; q (1-q) (1- 2\, q)  \; , \nonumber \\
  B_4(q) &=& -\frac{1}{30} + q^2(1-q)^2 \; .
  \label{bernoulli_defs}
\end{eqnarray}
These are valid only for $0 \leq q \leq 1$, and satisfy
\begin{equation}
  \frac{d }{dq} B_n(q) = n \, B_{n-1}(q) \; .
  \label{der_bernoulli}
\end{equation}

The effective potential in Eq.~(\ref{mass_action}) is proportional to
\begin{equation}
\int\frac{d^4K}{(2\pi)^4} \frac{1}{(K+Q)^2+m^2}=\frac{T^2}{12}\mathcal{A}(Q,m^2) \; ,
\end{equation}
where we introduce the shorthand notation,
\begin{equation}
  \int\frac{d^4K}{(2\pi)^4} =  T\sum_{n=-\infty}^\infty \int\frac{d^3K}{(2\pi)^3} \;,
\end{equation}
with   $K^\mu = (k_0, \vec{k})$,  $Q^\mu=(Q,\vec{0})$.  
The integral is evaluated as
\begin{equation}
  \mathcal{A}(Q,m^2)= \frac{12}{T^2}\int \frac{d^3k}{(2\pi)^3} \frac{1}{2E_k} \left( n(E_k-iQ)+n(E_k+iQ) \right) \; ,
  \label{definea}
\end{equation}
where $E_k=\sqrt{\vec{k}^2+m^2}$.
At nonzero holonomy the statistical distribution functions are
\begin{equation}
  n(E_k \mp iQ) = \frac{1}{ {\rm e}^{E_k/T \, \mp \, 2 \pi i \, q} -1 } \; .
\end{equation}
Thus $\mathcal{A}(Q,m^2)$ is manifestly periodic under $q \rightarrow q + 1$.
In the massless limit,
\begin{equation}
  \mathcal{A}(Q,0) = 6 \, B_2(|q|_{\rm mod \, 1})\; .
  \label{defA}
\end{equation}
From Eq.~(\ref{definea}), $\mathcal{A}(Q,m^2)$ is even in $Q$.
Along with periodicity, this implies that it is a function of the absolute value of $q$ modulo one, $|q|_{\rm mod \, 1}$.  

The potential is
\begin{equation}
\mathcal{V}_\text{m}
=C\sum_{a,b=1}^{N}\mathcal{P}^{a b}_{a b}\int \frac{d^4K}{(2\pi)^4}\frac{1}{(K+Q^{ab})^2}
=C \, \frac{T^2}{12} \sum_{a,b=1}^{N}\mathcal{P}^{a b}_{a b} \; \mathcal{A}(Q^{ab},0) \; ,
\label{eff_lag_massive}
\end{equation}
$Q^{ab}=Q^{a}-Q^b = 2 \pi T q^{ab}$, and $q^{a b}= q^a-q^b$.
The diagonal elements of the projector, $\mathcal{P}^{ab}_{ab} = 1 - \delta^{ab}/N$, enter to
ensure that the free energy is that for $SU(N)$ and not $U(N)$.

To this we add the perturbation contribution to the holonomous potential
\cite{Dumitru:2010mj,Dumitru:2012fw,Dumitru:2013xna,Guo:2014zra,Guo:2018scp,Pisarski:2012bj,Nishimura:2017crr,KorthalsAltes:2019yih,KorthalsAltes:2020ryu}.  The total effective potential is then
\begin{equation}
  \mathcal{V}(q) = \sum_{a,b=1}^N \mathcal{P}^{a b}_{a b}\; \left(
  \frac{2 \pi^2 T^4}{3} B_4(|q^{a b}|_{\rm mod \, 1}) + 
  \frac{C \, T^2}{2} \, B_2(|q^{a b}|_{\rm mod \, 1})  \right)
\; .
\end{equation}
By Eq.~(\ref{der_bernoulli}), the equations of motion involve $B_3$ and $B_1$.  For odd $n$,
$B_n(q)$ is periodic for positive $q$, but odd under $q \rightarrow -q$.  Thus the equation of motions are
\begin{equation}
  \sum_{b=1}^N \;
  {\rm sign}(q^{a b})
  \left( \frac{8 \pi^2 T^2}{3} \; B_3(|q^{ab}|_{\rm mod \, 1}) + C \,  B_1(|q^{ab}|_{\rm mod \, 1})\right) = 0 \; .
  \label{eom_massive}
\end{equation}
For two colors, the eigenvalues are $q$ and $-q$, with $|q^{1 2}| = 2 |q|$.  At small, positive $q$,
\begin{equation}
  \Bigl(4C+\frac{16 \pi^2 T^2}{3}\Bigr) \; q - \, C  = 0 \; .
  \label{eom_small_q}
\end{equation}
Thus at high $T \gg \sqrt{C}$,  $q \sim C/T^2$, and the holonomy is nonzero for any finite $T$.

For higher $N$, it is necessary to solve for the independent $q^a$'s, which are $N/2$ for even
$N$, and $(N-1)/2$ for odd $N \geq 3$.  
This can be done in the limit of large $N$ \cite{Dumitru:2012fw,Pisarski:2012bj}.
Nevertheless, it is clear that when $C \neq 0$, that $q^a$ is always nonzero, just because
the second Bernoulli polynomial starts out linear in $q^a$.

We {\it only} consider $q^a$'s which satisfy the equations of motion, and find
unexpected cancellations in the gluon self-energy.  The necessity of only looking at solutions
which satisfy the equations of motion was also found in studies of the free energy in the presence
of external sources \cite{KorthalsAltes:2019yih,KorthalsAltes:2020ryu}.

\subsection{Holonomous color current}
Consider the one-point function that contributes the expectation value of the color current $\langle J^{ab,\mu} \rangle$:
\begin{equation}
  \int\frac{d^4K}{(2\pi)^4}\frac{k_0+Q}{(K+Q)^2+m^2} =\frac{\pi T^3}{3} \; \mathcal{A}_0(Q,m^2) \; .
\end{equation}
At $Q = 0$, the integral vanishes automatically, as then the integrand is odd in $k_0$.  
It is nonzero when $Q \neq 0$,
\begin{equation}
  \mathcal{A}_0(Q,m^2) =\frac{3}{2i\pi T^3} \int \frac{d^3k}{(2\pi)^3} \bigl(n(E_k-iQ)-n(E_k+iQ)\bigr) \; .
  \label{defineA0}
\end{equation}
For a massless field,
\begin{equation}
  \mathcal{A}_0(Q,0) = 2\, {\rm sign}(q) \; B_3\left(|q|_{\rm mod \, 1} \right) \; .
  \label{a_b3}
\end{equation}
From its definition in Eq.~(\ref{defineA0}), $\mathcal{A}_0(Q,m^2)$ is odd in $Q$, which accounts
for the overall factor of ${\rm sign}(q)$ on the right hand side.

A simple trick can be used
to evaluate the derivatives of statistical distribution functions with respect to $m^2$ at $m^2=0$.
The mass only appears in the energy as $E_k=\sqrt{k^2+m^2}$, and so
a derivative of $E_k$ with respect to $m^2$ can be replaced by one with respect to $k^2$.
After that, it is direct to integrate by parts.  For example,
\begin{equation}
\frac{\partial}{\partial m^2} \int \frac{d^3 k}{(2 \pi)^3} \; n(E_k - i Q) = 
\int \frac{d^3 k}{(2 \pi)^3} \frac{\partial}{\partial k^2} n(E_k - i Q) =
- \frac{1}{4 \pi^2} \int_0^\infty dk \; n(E_k - i Q)  \; .
\label{trick}
\end{equation}
In this way, 
\begin{equation}
  \mathcal{A}'_0(Q)\equiv\left.\frac{\partial}{\partial m^2}\mathcal{A}_0(Q,m^2)\right|_{m^2=0} = 
  \frac{3}{(2 \pi T)^2} \; {\rm sign}(q) B_1\left(|q|_{\rm mod \, 1}\right) \; .
  \label{aprime_b1}
\end{equation}
Like $\mathcal{A}_0(Q)$, $\mathcal{A}'_0(Q)$ is odd in $Q$, which accounts for the overall factor of ${\rm sign}(q)$.

The expectation value of the color current has two contributions.   One is perturbative~\cite{Hidaka:2009hs},
\begin{equation}
    \begin{split}
      \langle J^{ab;\mu} \rangle_{\text{pt}}&=-2i g{f^{ab,cd,ef}}
      \mathcal{P}_{cd,ef}\int\frac{d^4K}{(2\pi)^4}\frac{(K^{cd})^\mu}{(K^{cd})^2}\\
    &=-u^\mu \delta^{ab}\frac{4\pi g T^3}{3\sqrt{2}}\sum_{c=1}^N\mathcal{A}_0(Q^{ac}) \; ,
  \end{split}
  \label{pert_color_courrent}
    \end{equation}
where $u^\mu = \delta^{\mu0}$.

The second contribution is from the auxiliary massive field,
\begin{equation}
   \begin{split}
     \langle J^{ab;\mu} \rangle_{\text{m}}&=
     -2ig{f^{ab,cd,ef}}\mathcal{P}_{cd,ef}C\frac{\partial}{\partial m^{2}}
     \left.\int\frac{d^4K}{(2\pi)^4}\frac{(K^{cd})^\mu}{(K^{cd})^2+m^{2}}\right|_{m^{2}=0}\\
   &
   =-u^{\mu}\delta^{ab}\frac{4\pi g T^3}{3\sqrt{2}}C\sum_{c=1}^N \mathcal{A}'_{0}(Q^{ac}) \;.
   \end{split}
   \end{equation}

The sum of the two contributions is
\begin{equation}
    \begin{split}
    \langle J^{ab;\mu} \rangle_{\text{total}}&=\langle J^{ab;\mu} \rangle_{\text{pt}}+\langle J^{ab;\mu} \rangle_{\text{m}}\\
    &=-u^\mu \delta^{ab}\frac{4\pi g T^3}{3\sqrt{2}}
    \sum_{c=1}^N\Bigl(\mathcal{A}_0(Q^{ac})+C\mathcal{A}'_{0}(Q^{ac})\Bigr)
    = 0 \;.
  \end{split}
  \label{a_eom}
   \end{equation}
From Eqs.~(\ref{a_b3}) and (\ref{aprime_b1}), this is equivalent to the equations of motion in Eq.~(\ref{eom_massive})
and so vanishes.

It is natural that the total color current vanishes in a consistent theory.  This also occurs
when the holonomous potential is computed in the presence of external, gauge invariant sources
\cite{KorthalsAltes:2019yih,KorthalsAltes:2020ryu}.
What is less obvious so is the computation of the gluon self-energies at nonzero holonomy,
to which we now turn.
   
\subsection{HHTL in the gluon self-energy}

We compute the gluon self-energy in the hard thermal loop (HTL) approximation
\cite{Hidaka:2008dr,Hidaka:2009hs,Hidaka:2009xh,Hidaka:2009ma,Bellac:2011kqa}.
We note that while we include the effect of the auxiliary massive field on the gluon propagator to $\sim g^2$,
we do {\it not} include the effect of the self-energy for the massive field.  Thus our analysis should only
be taken as a preliminary step towards a fully consistent effective theory.

Nevertheless, we show in Sec. (\ref{sec:hhtl}) that the effective gluon propagator with an auxiliary massive
field, or the two dimensional ghost of Sec. (\ref{sec:two_dim}), solves an important consistency check for the free
energy of a holonomous plasma, computed to $\sim g^3$ \cite{KorthalsAltes:2019yih,KorthalsAltes:2020ryu}.

In the Euclidean theory, the external momentum is $P_\mu^{12} = (p_0^{12},\vec{p})$,
where $p_0^{12} = p_0 + Q^1 + Q^2 = 2 \pi T (m + q^1 + q^2)$, for an integer $m$.
The HTL approximation requires that the external momenta are $\sim g T$, small relative to the temperature,
\begin{equation}
  |\vec{p}| \sim g T \ll T \;\; , \;\; |p_0^{12}| \sim g T \ll T \; .
\end{equation}
For the spatial momenta this is direct to implement.  For the timelike component, it is also
direct after analytic continuation to real energies, $p_0^{12} \rightarrow i \omega$,
as then we can directly let $\omega$, which is a continuous variable, be soft.
We computed the perturbative contribution previously in Ref. \cite{Hidaka:2009hs}.  

At nonzero holonomy, though, we can also compute a HTL
for Euclidean momenta.  We must work in the static limit, $p_0 = 0$, as otherwise the energy
$p_0$ is $2 \pi T$ times some nonzero integer, $m$.  At zero holonomy this is just the static limit.
At nonzero holonomy, however, we obtain a nontrivial limit simply
by requiring that $Q_1$ and $Q_2$ are soft, $\sim g T$.

In both cases we term the result a holonomous hard thermal loop, or HHTL.   For real energies, the HHTL in the gluon
self-energy is $g^2 T^2$ times a function of the dimensionless variable, $\omega/p$, and $\hat{p}$.
In the Euclidean
theory, there is an analogous gluon self-energy which is $g^2 T^2$ times a function of the dimensionless
variable, $Q/p$, and $\hat{p}^i$.  In both cases, the HHTL is important because for soft momenta, the
inverse propagator is $\sim P^2 \sim g^2 T^2$ (modulo singularities), and the HHTL is as large as the
term at tree level.  Thus the HHTL must be included self-consistently in order to compute for soft momenta.

The computation of the contribution of the massive field to the HHTL in the gluon self-energy
is a straightforward generalization of the usual perturbative computation \cite{Hidaka:2009hs}.  
For a light but massive field, the HHTL is $g^2$ times the loop integral \cite{Hidaka:2009hs}
\begin{equation}
  \frac{\partial}{\partial m^2}\left.\widetilde{\mathcal{J}}^{\mu\nu}(P^{12},Q_1,Q_2,m^2)\right|_{m^2=0} \; ,
\end{equation} 
where
\begin{equation}
  \widetilde{\mathcal{J}}^{\mu\nu}(P^{12},Q_1,Q_2,m^2)=\mathcal{J}^{\mu\nu}(P^{12},Q_1,Q_2,m^2)
  -\frac{\delta^{\mu\nu}}{4}\frac{T^2}{12}(\mathcal{A}(Q_1,m^2)+\mathcal{A}(Q_2,m^2) ) \; ,
  \label{htl_general}
\end{equation}
and
\begin{equation}
  \mathcal{J}^{\mu\nu}(P^{12},Q_1,Q_2,m^2)=
  \frac{1}{8}T \sum_{n=-\infty}^{+\infty}
  \int\frac{d^3k}{(2\pi)^3}\frac{(2K^1-P^{12})^\mu (2K^1-P^{12})^\nu}{((K^1)^2+m^2)((P^{12}-K^1)^2+m^2)}+(Q_1\leftrightarrow Q_2) \; .
\end{equation}
The loop $k_0 = 2 \pi T n$ for integral $n$, while $K^1=K+Q^1$, with $k_0^1 = k_0 + Q^1 = 2 \pi T (n + q^1)$.  

The sum over the Matsubara frequency $n$ can be done by going to a coordinate representation in the Euclidean
time, $\tau$ \cite{Hidaka:2009hs}.  The result is simplest for the spatial components of
$\mathcal{J}^{\mu\nu}$,
\begin{equation}
  \mathcal{J}^{ij}(P^{12},Q_1,Q_2,m^2)=\frac{1}{8}\int\frac{d^3k}{(2\pi)^3}
  \frac{(2k-p)^i(2k-p)^j}{(2E_k)(2E_{p-k})}(\mathcal{I}_1+\mathcal{I}_2+\mathcal{I}_3+\mathcal{I}_4)+(Q_1\leftrightarrow Q_2)\; ,
  \end{equation}
  where
  \begin{align}
    \mathcal{I}_1&=\frac{-1}{ip^{12}_0-E_k-E_{p-k}}\Bigl(1+n(E_k-iQ_1)+n(E_{p-k}-iQ_2)\Bigr)\; ,\\
    \mathcal{I}_2&=\frac{1}{ip^{12}_0-E_k+E_{p-k}}\Bigl(n(E_k-iQ_1)-n(E_{p-k}+iQ_2)\Bigr)\; ,\label{landau2}\\
    \mathcal{I}_3&=\frac{-1}{ip^{12}_0+E_k-E_{p-k}}\Bigl(n(E_k+iQ_1)-n(E_{p-k}-iQ_2)\Bigr)\; ,\label{landau3}\\
    \mathcal{I}_4&=\frac{1}{ip^{12}_0+E_k+E_{p-k}}\Bigl(1+n(E_k+iQ_1)+n(E_{p-k}+iQ_2)\Bigr) \; .
  \end{align}
  The easiest terms to evaluate are those $\sim \mathcal{I}_1$ and $\sim \mathcal{I}_4$,
  as they do not involve Landau damping.  In this case, the external momentum 
$P^{12}$ can be neglected, and these terms reduce to
\begin{equation}
    \begin{split}
    &\frac{1}{8}\int\frac{d^3k}{(2\pi)^3}\frac{(2k-p)^i(2k-p)^j}{(2E_k)(2E_{p-k})}(\mathcal{I}_1+\mathcal{I}_4)+(Q_1\leftrightarrow Q_2)\\
    &\stackrel{\text{HTL}}\approx
    \frac{\delta^{ij}}{12}\int\frac{d^3k}{(2\pi)^3}
    \frac{\vec{v}_k^2}{2E_k}\Bigl(n(E_k-iQ_1)+n(E_k+iQ_1)+n(E_{k}-iQ_2)+n(E_{k}+iQ_2)\Bigr)\; ,
  \end{split}
  \label{I14_HTL}
  \end{equation}
where we introduce the vectors $\vec{v}_k= {\vec{k}}/{E_k}$.

  More care must be taken with the terms $\sim \mathcal{I}_2$ and $\sim \mathcal{I}_3$, as they involve Landau
  damping, and diverge as $P \rightarrow 0$.  Expanding to terms linear in $P$,
\begin{equation}
  \begin{split}
    E _{p - k}  
    &\cong E_{k} - \vec{v}_{k}\cdot\vec{p} \; ,\\ 
    n (E_{p - k}-iQ )& \cong n (E_{k}-iQ )- (\vec{v}_{k}\cdot\vec{p}) \; n'(E_k -iQ) \; ,\\
    ip^{12}_0  + E_k- E_{p - k}    &\cong ip_0^{12} + \vec{v}_{k}\cdot\vec{p}= P^{12} \cdot \widetilde{K} \; ,
  \end{split}
\end{equation}
where
\begin{equation}
  n' (E_k -iQ)=\frac{\partial }{\partial E_k } \; n (E_k -iQ) \; ,
  \label{derv_n}
\end{equation}
and $ \widetilde{K}=(i,{\vec{v}}_{k})$.
For future reference, we note that for massless fields $E_{k} \rightarrow |\vec{k}|=k$, and these vectors become
\begin{equation}
\vec{v}_{k} \rightarrow \widehat{\vec{k}}\;\; , \;\;  \widetilde{K} \rightarrow \widehat{K}=(i,\widehat{\vec{k}}) \; .
\label{define_hatK}
\end{equation}
For massive fields, $\widetilde{K}^2 = - m^2/E_k^2$, while $\widehat{K}^\mu$ is null, $\widehat{K}^2 = 0$.

In the HTL approximation, $\mathcal{I}_2$ and $\mathcal{I}_3$ become
\begin{equation}
  \begin{split}
    &\frac{1}{8} \int\frac{d^3k}{(2\pi)^3}
    \frac{(2k-p)^i(2k-p)^j}{(2E_k)(2E_{p-k})}(\mathcal{I}_2+\mathcal{I}_3)+(Q_1\leftrightarrow Q_2)\\
    &\stackrel{\text{HTL}}\approx \frac{1}{4}\int\frac{d^3k}{(2\pi)^3}\; v_k^iv_k^j \; \Bigl[
      \frac{1}{P^{12}\cdot\widetilde{K}}\Bigl(n(E_k-iQ_1)-n(E_k+iQ_1)
      +n (E_{k}-iQ_2 )-n (E_{k}+iQ_2 )\Bigr)\\&
      \;\;\;\;\;\;\;\;\;\;\;\;\;\;\;\; - \frac{1}{2}\frac{\vec{v}_{k}\cdot\vec{p}}{P^{12}\cdot\widetilde{K}}
      \Bigl( n' (E_k -iQ_1)+ n' (E_k +iQ_1)+n' (E_k -iQ_2)+ n' (E_k +iQ_2) \Bigr)
    \Bigr] \;.
  \end{split}
\end{equation}
The second line can be rewritten as
\begin{equation}
  \begin{split}
    &-\frac{1}{8}\int\frac{d^3k}{(2\pi)^3}v_k^iv_k^j
    \frac{\vec{v}_{k}\cdot\vec{p}}{P^{12}\cdot\widetilde{K}}
    \Bigl( n' (E_k -iQ_1)+ n' (E_k +iQ_1)+n' (E_k -iQ_2)+ n' (E_k +iQ_2) \Bigr)\\
      &=\frac{1}{8}\int\frac{d^3k}{(2\pi)^3}\Bigl[  
      ip_0^{12}\, \frac{v_k^iv_k^j}{P^{12}\cdot\widetilde{K}}
      \Bigl( n' (E_k -iQ_1)+ n' (E_k +iQ_1)+n' (E_k -iQ_2)+ n' (E_k +iQ_2) \Bigr)\\
      &\;\;\;\;\;\;\;\;\;\;\;\;\;\;\;\;
      -v_k^iv_k^j\Bigl( n' (E_k -iQ_1)+ n' (E_k +iQ_1)+n' (E_k -iQ_2)+ n' (E_k +iQ_2) \Bigr)
          \Bigr]\\
      &=\frac{1}{8}\int\frac{d^3k}{(2\pi)^3}\Bigl[  
      ip_0^{12} \, \frac{v_k^iv_k^j}{P^{12}\cdot\widetilde{K}}
      \Bigl( n' (E_k -iQ_1)+ n' (E_k +iQ_1)+n' (E_k -iQ_2)+ n' (E_k +iQ_2) \Bigr)\\
      &\;\;\;\;\;\;\;\;\;\;\;\;\;\;\;\;
      +\delta^{ij}\Bigl(\frac{1}{E_k}-\frac{\vec{v}_k^2}{3E_k}\Bigr)
      \Bigl( n (E_k -iQ_1)+ n (E_k +iQ_1)+n (E_k -iQ_2)+ n (E_k +iQ_2) \Bigr)
           \Bigr] \;.
        \end{split}
\end{equation}
In the last line, we replaced $v_k^iv_k^j\to \delta^{ij}\vec{v}_k^2/3$ and then integrated by parts.
Collecting these results, we find
\begin{equation}
  \begin{split}
  &\widetilde{\mathcal{J}}^{ij}(P^{12},Q_1,Q_2,m^2)\\
     &=\frac{1}{4}\int\frac{d^3k}{(2\pi)^3}\Bigl[ 
    \Bigl[
      \frac{v_k^i v_k^j }{P^{12}\cdot\widetilde{K}}\; \Bigl(n(E_k-iQ_1)-n(E_k+iQ_1)
      +n (E_{k}-iQ_2)-n (E_{k}+iQ_2 ) \Bigr)
  \\
  &\quad+ ip_0^{12} \, \frac{v_k^i v_k^j}{P^{12}\cdot\widetilde{K}} \; \frac{1}{2}
  \Bigl( n' (E_k -iQ_1)+ n' (E_k +iQ_1)+n' (E_k -iQ_2)+ n' (E_k +iQ_2) \Bigr)
       \Bigr] \;.
  \end{split}
\end{equation}
$\widetilde{\mathcal{J}}^{0j}$ and $\widetilde{\mathcal{J}}^{00}$ follow from the relation
\begin{equation}
  \begin{split}
  P^{12}_\mu\widetilde{\mathcal{J}}^{\mu\nu}(P^{12},Q_1,Q_2,m^2)
  &=-\frac{1}{2}\int\frac{d^4K}{(2\pi)^4}\frac{ (K^2)^\nu}{(K^2)^2+m^2}
  -\frac{1}{2}\int\frac{d^4K}{(2\pi)^4}\frac{ (K^{1})^\nu}{(K^1)^2+m^2}\\
  &=-\frac{u^\nu}{2}\frac{\pi T^3}{3}(\mathcal{A}_0(Q_1,m^2)+\mathcal{A}_0(Q_2,m^2) ) \;.
  \end{split}
\end{equation}
The final result for $\widetilde{\mathcal{J}}^{\mu\nu}(P^{12},Q_1,Q_2,m^2)$ is
\begin{equation}
  \begin{split}
  &\widetilde{\mathcal{J}}^{\mu\nu}(P^{12},Q_1,Q_2,m^2)\\
  &=\frac{1}{4}\int\frac{d^3k}{(2\pi)^3}
    \Bigl[
      \frac{\widetilde{K}^\mu \widetilde{K}^\nu}{P^{12}\cdot\widetilde{K}}\Bigl(n(E_k-iQ_1)-n(E_k+iQ_1)
      +n (E_{k}-iQ_2 )-n (E_{k}+iQ_2 )\Bigr)
  \\
  &\quad+\Bigl(u^\mu u^\nu+ ip_0^{12} \; \frac{\widetilde{K}^\mu\widetilde{K}^\nu}{P^{12}\cdot\widetilde{K}}\Bigr)
  \frac{1}{2}\Bigl( n' (E_k -iQ_1)+ n' (E_k +iQ_1)+n' (E_k -iQ_2)+ n' (E_k +iQ_2) \Bigr)
       \Bigr] \;.
     \end{split}
     \label{final_massive}
\end{equation}
This is the HTL approximation for a massive particle at nonzero holonomy, $Q \neq 0$.  Notice that the terms
$\sim \delta^{i j}$ have cancelled between Eqs.~(\ref{htl_general}) and (\ref{I14_HTL}).

In the massless limit this reduces to the usual HTL loops~\cite{Hidaka:2009hs}:
\begin{equation}
  \widetilde{\mathcal{J}}^{\mu\nu}(P^{12},Q_1,Q_2,0)
  \stackrel{\text{HTL}}\approx
  \frac{i\pi T^3}{6}(\mathcal{A}_0(Q_1)+\mathcal{A}_0(Q_2))\delta\Gamma^{\mu\nu}(P^{12})
  +\frac{T^2}{24}(\mathcal{A}(Q_{1})+\mathcal{A}(Q_{2}))\delta\Pi^{\mu\nu}(P^{12}) \; ,
\end{equation}
where
\begin{align}
  \delta\Gamma^{\mu\nu}(P)&=\int\frac{d\Omega}{4\pi}\; 
                               \frac{\widehat{K}^\mu \widehat{K}^\nu}{P\cdot\widehat{K}}\label{delta_gamma}\; ,\\
  \delta\Pi^{\mu\nu}(P)&= - u^\mu u^\nu - ip_0\int
                            \frac{d\Omega}{4\pi}\frac{\widehat{K}^\mu\widehat{K}^\nu}{P\cdot\widehat{K}} \; ,
                            \label{landauQ_T}
\end{align}
$d\Omega = d\cos\theta d\phi$ with the polar angle $\theta$ and the azimuthal angle $\phi$, and
$\widehat{K}$ is defined in Eq.~(\ref{define_hatK}).

These functions satisfy
\begin{equation}
  P_\mu\; \delta\Gamma^{\mu\nu}(P)=iu^\nu
  \;\; , \;\; P_\mu \; \delta\Pi^{\mu\nu}(P)=0 \; .
  \label{pi_transverse}
\end{equation}
Because $\Gamma^{\mu \nu}(P)$ is not transverse, neither is this part of the gluon self-energy.
We discuss at the end of this section how transversity is restored when all contributions are included.

For the massive field, we need the derivative of $ \widetilde{\mathcal{J}}^{\mu\nu}$ with respect to $m^2$,
evaluated at $m^2 = 0$.  The mass appears only through the energy $E_k=\sqrt{k^2+m^2}$, which
appears in terms involving $\widetilde{K}^i = \vec{k}^i/E_k$
and the statistical distribution functions, such as $n(E_k-Q_1)$.  Note that $\widetilde{K}^0 = i$ is independent
of the energy and so the mass.

We require
\begin{equation}
\begin{split}
\frac{\partial}{\partial m^{2}} \frac{\widetilde{K}^{\mu}\widetilde{K}^{\nu}}{P^{12}\cdot \widetilde{K}}
=\frac{1}{P^{12}\cdot \widetilde{K}}\; \frac{\partial}{\partial m^{2}}\widetilde{K}^{\mu}\widetilde{K}^{\nu}
-\; \frac{\widetilde{K}^{\mu}\widetilde{K}^{\nu}}{(P^{12}\cdot \widetilde{K})^{2}}\;
\frac{\partial}{\partial m^{2}}P^{12}\cdot \widetilde{K} \; .
\end{split}
\end{equation}
As
\begin{align}
\frac{\partial}{\partial m^{2}}P^{12}\cdot \widetilde{K}&=
-\vec{p}\cdot\vec{v}_{k}\frac{1}{2E_{k}^{2}}=\frac{1}{2E_{k}^{2}}(ip_0^{12}-P^{12}\cdot\widetilde{K}) \;,\\
\frac{\partial}{\partial m^{2}}\widetilde{K}^{\mu}\widetilde{K}^{\nu}&=
-\frac{1}{2E_{k}^{2}} (2\widetilde{K}^{\mu}\widetilde{K}^{\nu}-iu^{\mu}\widetilde{K}^{\nu}-iu^{\nu}\widetilde{K}^{\mu})
\; ,
\end{align}
we find
\begin{equation}
\begin{split}
\left.\frac{\partial}{\partial m^{2}}\frac{\widetilde{K}^{\mu}\widetilde{K}^{\nu}}{P^{12}\cdot \widetilde{K}}\right|_{m^2=0}
&=\frac{-1}{2k^{2}}\Bigl(
\frac{1}{P^{12}\cdot \widehat{K}}
(\widehat{K}^{\mu}\widehat{K}^{\nu}-iu^{\mu}\widehat{K}^{\nu}-iu^{\nu}\widehat{K}^{\mu})
+\frac{\widehat{K}^{\mu}\widehat{K}^{\nu}}{(P^{12}\cdot \widehat{K})^{2}}  ip_0^{12}
\Bigr) \; .
\end{split}
\label{derivative_mass}
\end{equation}

For the massive fields, the momentum integrals which arise include those of Eq.~(\ref{aprime_b1}). 
Similarly, using the trick of Eq.~(\ref{trick}), 
\begin{equation}
\begin{split}
  &\frac{\partial}{\partial m^2} \left.
    \int\frac{d^{3}k}{(2\pi)^{3}}\frac{1}{2}
    \Bigl( n' (E_k -iQ_1)+ n' (E_k +iQ_1)+n' (E_k -iQ_2)+ n' (E_k +iQ_2) \Bigr)\right|_{m^2=0}\\
  &=- \frac{1}{8\pi^{2}}\int_0^\infty dk \Big(n'(k-iQ_1)+  n'(k+iQ_1)+ n'(k-iQ_2) 
    + n'(k+iQ_2)\Big)=-\frac{1}{4\pi^{2}} \; .\\
\end{split}
\label{der_mass_der_ns}
\end{equation}
Here we have used a peculiar identity at zero energy and nonzero holonomy,
\begin{equation}
  n(i Q) + n(- i Q) =   \frac{1}{{\rm e}^{+ i Q/T} - 1} + \frac{1}{{\rm e}^{- i Q/T} - 1}  = -1 \; .
  \label{identity_sum}
\end{equation}
Substituting Eqs.~(\ref{aprime_b1}), ~\eqref{derivative_mass}, and \eqref{der_mass_der_ns}
into $\partial/\partial m^2$ of Eq.~\eqref{final_massive}, at $m^2 = 0$ we obtain
\begin{equation}
  \begin{split}
    \frac{\partial}{\partial m^2}\left.\widetilde{\mathcal{J}}^{\mu\nu}(P^{12},Q_1,Q_2,m^2)\right|_{m^2=0}
   &\stackrel{\text{HTL}}\approx
   \frac{i \pi T^3}{6}(\mathcal{A}_0'(Q_1)+\mathcal{A}_0'(Q_2))\delta\Gamma^{\mu\nu}(P^{12})
    +\frac{1}{16\pi^2}\delta\Pi^{\mu\nu}(P^{12})\\
    &\quad-\frac{1}{16\pi^2}\Bigl(p_0^{12}-\frac{8\pi^3 T^3}{3}
    (\mathcal{A}'_0(Q_1)+\mathcal{A}'_0(Q_2)) \Bigr)\delta\widetilde{\Pi}^{\mu\nu}(P^{12}) \; ,
      \end{split}
    \end{equation}
    where
\begin{equation}
  \delta\widetilde{\Pi}^{\mu\nu}(P)=\int\frac{d\Omega}{4\pi}\frac{i}{P\cdot \widehat{K}}\Bigl[
    (\widehat{K}^{\mu}-iu^\mu)(\widehat{K}^{\nu}-iu^\nu)
    +\Bigl(u^\mu u^\nu+ip_0\frac{\widehat{K}^{\mu}\widehat{K}^{\nu}}{P\cdot \widehat{K}}\Bigr)
    \Bigr] \; .
    \label{tilde_Pi}
\end{equation}
This tensor is transverse in the external momentum,
\begin{equation}
  P_\mu \; \delta\widetilde{\Pi}(P) = \int \frac{d\Omega}{4 \pi} \left( i\widehat{K}^\nu + u^\nu \right)
  = 0 \; ,
  \label{trans_tilde_pi}
\end{equation}
after performing the angular integral.

The perturbative contribution to the gluon self-energy was computed in Ref. \cite{Hidaka:2009hs},
\begin{equation}
  \Pi_{\text{pt}}^{ab,cd;\mu\nu}(P^{cd}) \stackrel{\text{HTL}}\approx -4g^2f^{(ab,ef,gh)}{f^{(cd,fe,hg)}}
  \widetilde{\mathcal{J}}^{\mu\nu}(P^{ab},Q^{fe},Q^{hg},0) \; .
\end{equation}
The analogous contribution of the massive field is
\begin{equation}
  \begin{split}
    \Pi_{\text{m}}^{ab,cd;\mu\nu}(P^{cd}) \stackrel{\text{HTL}}\approx
    -4\, g^2 C\, f^{(ab,ef,gh)}{f^{(cd,fe,hg)}}\frac{\partial}{\partial m^{2}}\left.
  \widetilde{\mathcal{J}}^{\mu\nu}(P^{ab},Q^{fe},Q^{hg},m^2)\right|_{m^{2}=0} \; .
  \end{split}
\end{equation}

\begin{figure}
\begin{center}
\begin{align*}
\parbox{3.8cm}{\includegraphics[width=.23\textwidth]{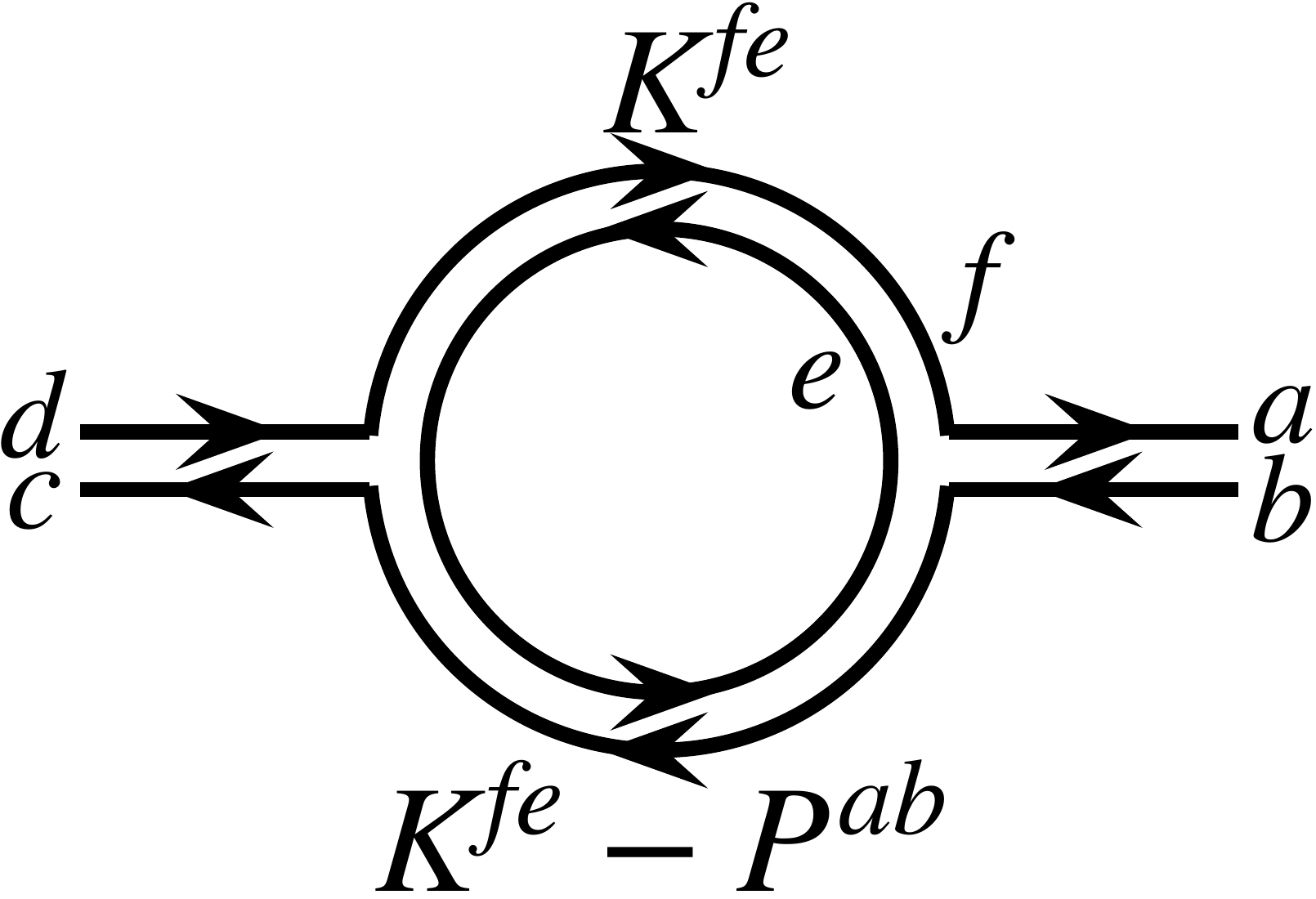}}
+\parbox{3.8cm}{\includegraphics[width=.23\textwidth]{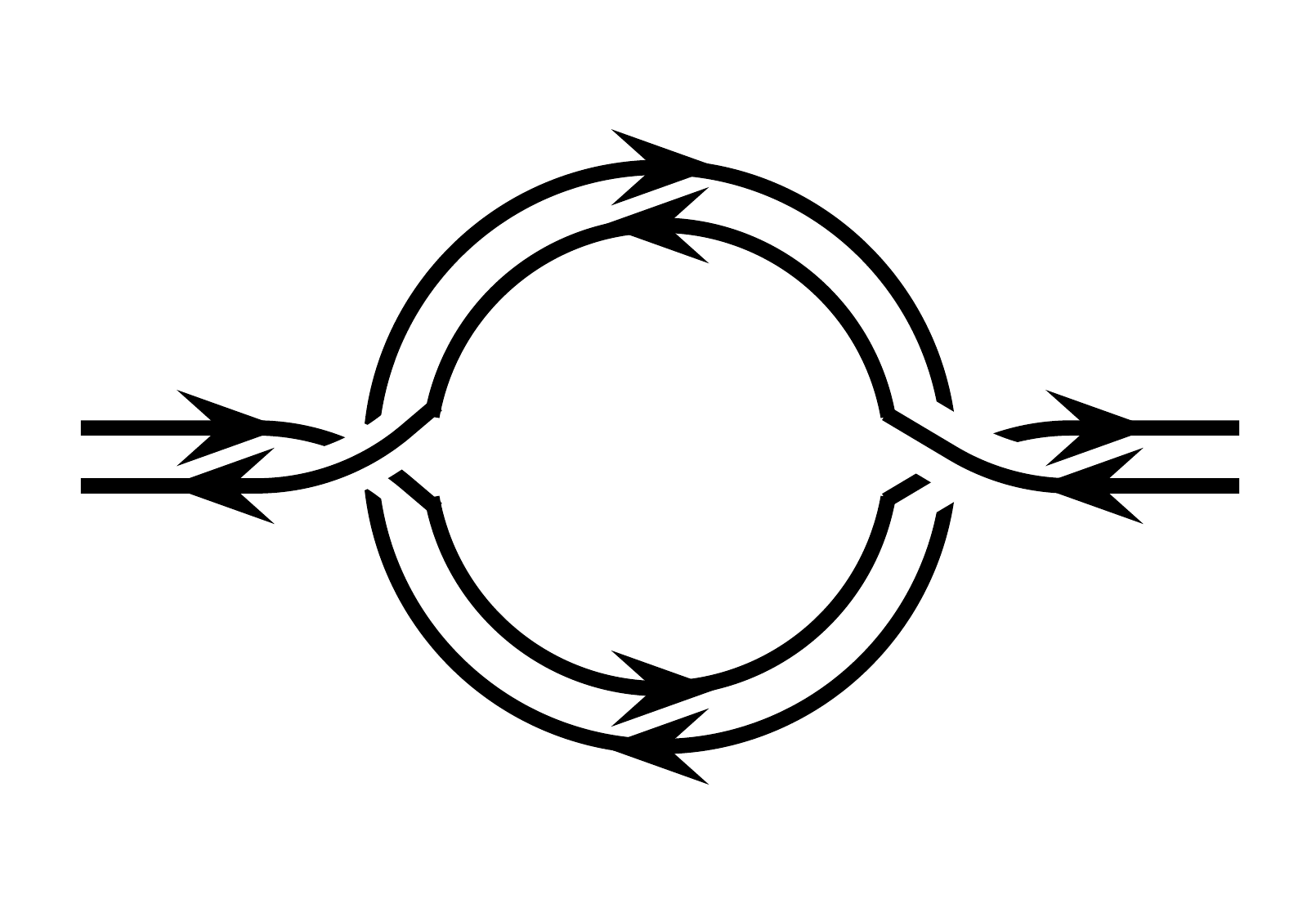}}
-\parbox{3.8cm}{\includegraphics[width=.23\textwidth]{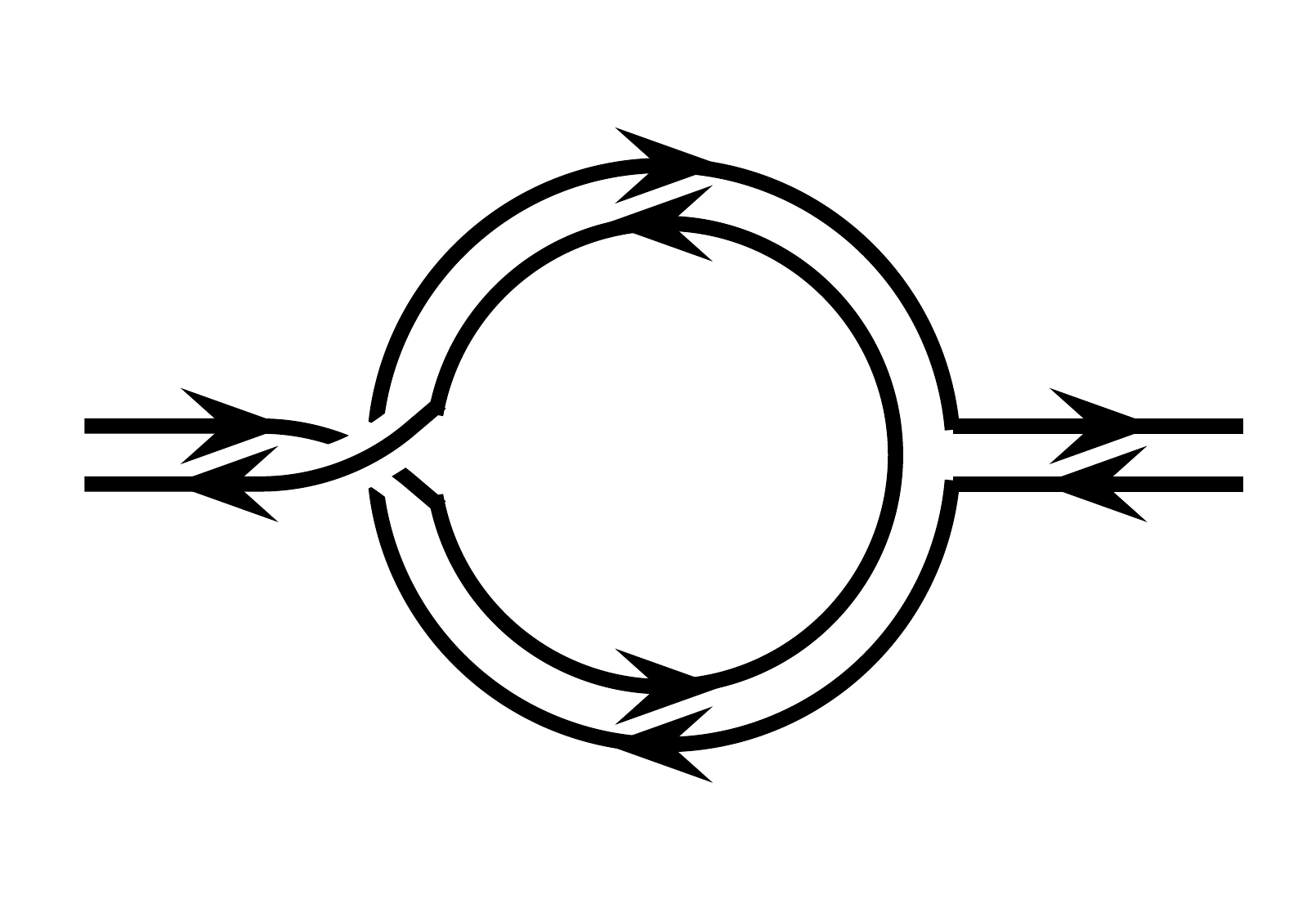}}
-\parbox{3.8cm}{\includegraphics[width=.23\textwidth]{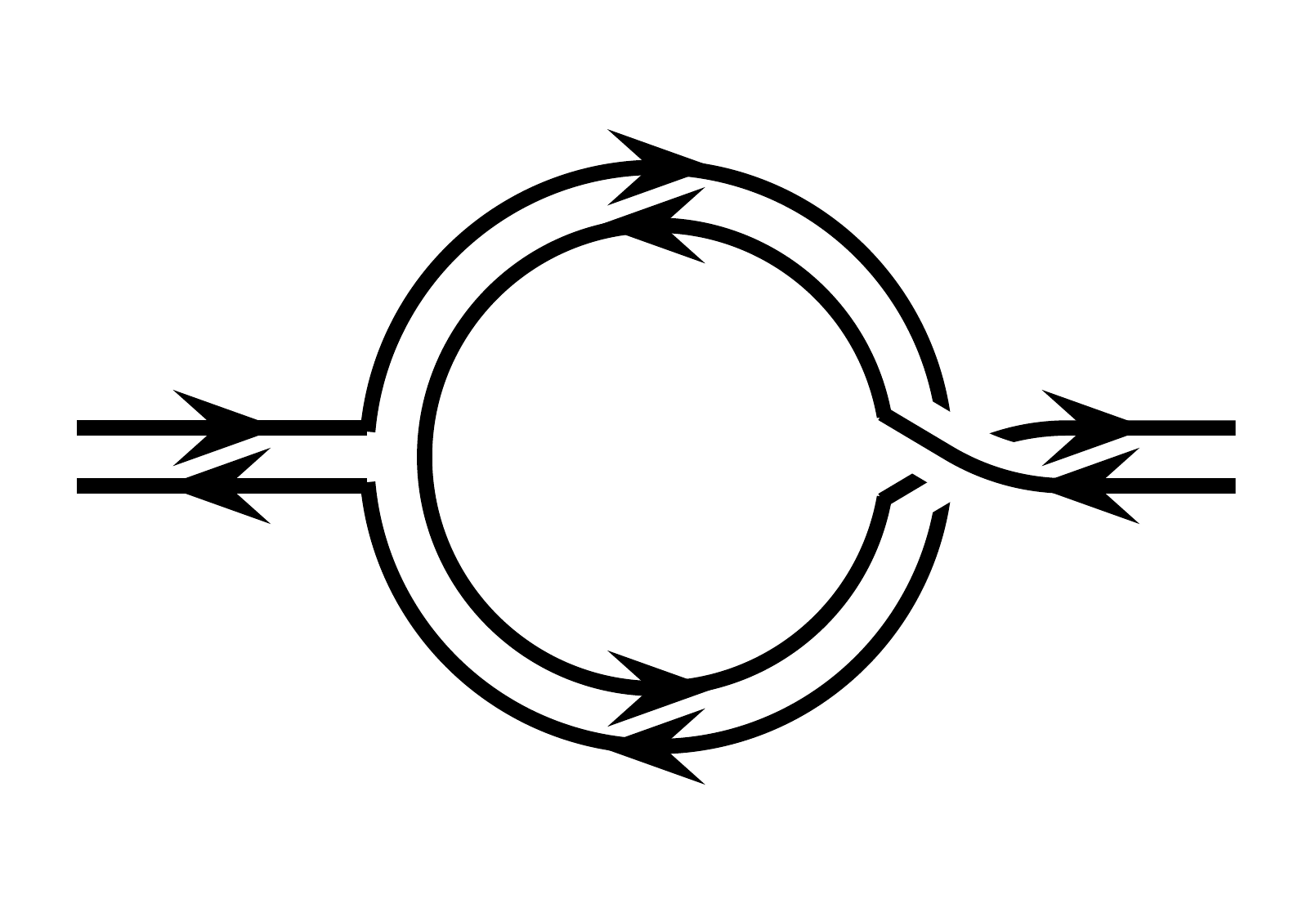}}
\end{align*}
\end{center}
\caption{Product of the structure functions, times $2$, which enter into the gluon self-energy at one-loop order.}
  \label{fig:one-loop_gluon}
\end{figure}
The product of the structure functions simplifies in the double
line notation, as illustrated in Fig. (\ref{fig:one-loop_gluon}).  The result is
\begin{equation}
  \begin{split}
    \Pi_{\text{total}}^{ab,cd;\mu\nu}(P^{cd})& =
    \Pi_{\text{pt}}^{ab,cd;\mu\nu}(P^{cd})+\Pi_{\text{m}}^{ab,cd;\mu\nu}(P^{cd}) \\
  &\stackrel{\text{HTL}}\approx  -{\cal K}^{ab,cd} \delta\Gamma^{\mu\nu}(P^{ab}) - \left(m^2_\text{gl}\right)^{ab,cd}
    \delta\Pi^{\mu\nu}(P^{ab}) - \left(\widetilde{m}^3_\text{gl}\right)^{ab,cd}\delta\widetilde{\Pi}^{\mu\nu}(P^{ab}) \; ,
  \end{split}
  \label{massive_self_energy}
\end{equation}
where
\begin{equation}
    {\cal K}^{ab,cd}(Q) = \frac{2i \pi g^2 T^3}{3} \delta^{ad}\delta^{bc} \sum_{e=1}^N \Bigl( \mathcal{A}_0(Q^{ae})+
    \mathcal{A}_0(Q^{eb})+C(\mathcal{A}_0'(Q^{ae})+\mathcal{A}_0'(Q^{eb}))\Bigr) = 0 \; ,
    \label{K_result}
  \end{equation}
  \begin{equation}
    (m^2_\text{gl})^{ab,cd}(Q) = \frac{g^2T^2}{6}\Bigl(
    \delta^{ad}\delta^{bc} \sum_{e=1}^N \left(\mathcal{A}(Q^{ae})+\mathcal{A}(Q^{eb})\right)
    - 2 \delta^{a b} \delta^{cd} \mathcal{A}(Q^{a c}) \Bigr)
    + g^2 \frac{C N}{4 \pi^2} P^{ab, cd} \; ,
    \label{m2_result}
  \end{equation}
  \begin{equation}
    (\widetilde{m}^3_\text{gl})^{ab,cd}(p_0,Q)
    = -\; \frac{g^2 CN}{4\pi^2} {\cal P}^{ab, cd}
    \; \Bigl(p_0^{ab}-\frac{8\pi^3 T^3}{3N}\sum_{e=1}^N (\mathcal{A}'_0(Q^{ae})+\mathcal{A}'_0(Q^{eb})) \Bigr) \; .
    \label{tildem2_result}
    \end{equation}

The first term, $\sim {\cal K}^{ab,cd}$, vanishes by the equations of motion, Eq.~(\ref{a_eom}).
We originally conjectured in Ref. \cite{Hidaka:2009hs} that terms which contribute to $\sim \mathcal{K}^{ab,cd}$
must cancel.  It is natural that they do so, because by Eq.~(\ref{a_eom}) the equations of motion are proportional
to the color current, and for any consistent theory nonzero holonomy
should not induce a net color current.  Nevertheless, it is
gratifying to see it emerge naturally.

This leaves the terms $\sim \delta\Pi^{\mu \nu}(P^{ab})$ and
$\sim \delta \widetilde{\Pi}^{\mu \nu}(P^{a b})$.  These two are both transverse in the external momentum, which is
necessary to ensure gauge invariance.
The term $m^2_\text{gl}$ is just the generalization of the Debye masses to nonzero holonomy, as
$\delta \Pi^{\mu \nu}$ is the standard hard thermal loop.
The term $\widetilde{m}^3_\text{gl}$ is special to taking a massive auxiliary field, and does not arise
for two dimensional ghosts.  Since $C$ has dimensions of mass squared, 
$\widetilde{m}^3_\text{gl}$ has dimensions of mass cubed.  

It is useful to contrast these results with the introduction of an external source.
Consider first a scalar field $\chi$.  To induce $\langle \chi \rangle \neq 0$,
one adds to the action a source linear in $\chi$, such as $\sim \int d^4 x \, J(x) \chi(x)$.  
This shifts the expectation value of $\chi$, but obviously doesn't 
affect any higher point function.

To induce nonzero holonomy for the gauge field, however, it is best to use a gauge invariant source,
which at nonzero temperature are sums over Polyakov loops;
for analyses with sources linear in $A_0$, see Refs.
\cite{Skalozub:1992un,Skalozub:1992pb,Skalozub:1994br,Skalozub:1994pa,Borisenko:1994jn,Skalozub:2004ab,Bordag:2018aii,Borisenko:2020dej,Skalozub:2020idf}.
For example,
we can add a source term which is a sum over squares of Polyakov loops,
as in Eq.~(71) of Ref. \cite{KorthalsAltes:2020ryu}:
\begin{equation}
  {\cal S}_\epsilon = \epsilon \int \frac{d^3 x }{V} \sum_{r = 1}^{r_{0}} c_r |{\rm tr} \; {\bf L}^r(x)|^2 \;\; , \;\;
  {\bf L}(x) = {\cal P} \; {\rm exp} \left( i g \int^{1/T}_0 \, A_0(x,\tau) d\tau \right) \; .
  \label{external_source}
\end{equation}
${\bf L}(x)$ is the thermal Wilson line in the imaginary time $\tau$ and the $c_r$ are arbitrary coefficients.
It is also possible to take a sum over linear powers of Polyakov loops, Eq.~(20) of Ref.
\cite{KorthalsAltes:2019yih}, as the gluon self-energy in
Eq.~(\ref{self_energy_source}) is unchanged.  The source term
${\cal S}_\epsilon$ induces nonzero holonomy when expanded 
to first order in quantum fluctuations, Eq.~(72) of Ref. \cite{KorthalsAltes:2020ryu}.  As a series of exponentials
in $A_0$, though, ${\cal S}_\epsilon$ induces
an {\it infinite} series of higher point functions for quantum fluctuations.
For the two point function and Euclidean momenta, the source contributes to the gluon self-energy as
\begin{equation}
  \Pi_{\epsilon}^{00; ab,cd}(P^{ab}) = -
  \frac{2 \pi g^2 T^3 }{3} \; \delta^{ad} \delta^{bc} \; \frac{1}{p_0^{a b}} \;
  \sum_{e=1}^N
  \left(
    {\cal A}_0\left(Q^a - Q^e\right)  + {\cal A}_0\left(Q^e - Q^b\right)
  \right) \; ,
  \label{self_energy_source}
\end{equation}
Eq.~(35) of Ref. \cite{KorthalsAltes:2019yih} and Eqs.~(51) and (74) of Ref. \cite{KorthalsAltes:2020ryu}.
This is derived by using the equations of motion for the $Q^a$'s,
which eliminates {\it any} dependence on $\epsilon$ and the coefficients $c_r$.

Computing only the perturbative contributions to the gluon self-energy, the gluon self-energy is not
transverse.  In the presence of an external source, the total gluon self-energy is a sum of the
perturbative and source terms, 
$\Pi_{\rm total, \epsilon}^{\mu\nu} = \Pi_{\rm pt}^{\mu\nu} + \Pi_\epsilon^{\mu\nu}$.
Then for all values of the gauge fixing parameter this total gluon self-energy satisfies:
\begin{equation}
Q^{a b}\Pi_{\rm total , \epsilon}^{0 \nu; ab,cd}(Q^{a b},0)= 0 \;\; , \;\;
P_\mu^{a b} P_\nu^{a b} \Pi_{\rm total , \epsilon}^{\mu \nu; ab,cd}(P^{a b})= 0 \; ,
\label{constraint_transverse}
\end{equation}
Eqs.~(68) and (70) of Ref. \cite{KorthalsAltes:2020ryu}.  The former holds for $p_0 = p = 0$ and $Q^{ab} \neq 0$,
the latter for all $P_\mu^{a b}$.  
The latter is necessary to establish gauge invariance in
computing the free energy to $\sim g^3$ at soft holonomy,
and to $\sim g^4$ and beyond for any $Q$ \cite{KorthalsAltes:2020ryu,altes_nishimura}.

In Ref. \cite{KorthalsAltes:2020ryu} 
it was also argued that in order to have a source for which the holonomy turns
on smoothly as $\epsilon$ increases, that one must sum over an {\it infinite} number of loops, $r_{0} = \infty$.
This was established carefully for $N=2$ and $\infty$, and is very reasonable for any $N$.
Indeed, it is very natural to take $c_r = 1/r^2$ and $r_0 = \infty$, so that ${\cal S}_\epsilon$ is proportional
to the second Bernoulli polynomial, $B_2(q)$.  As discussed in the Introduction,
Sec. (\ref{sec:introduction}), an infinitesimal value of the source generates
a corresponding holonomy which is also infinitesimal.  For a scalar field with source $\sim J \chi$, this is
trivial, but with a source of Polyakov loops, it is not trivial to ensure \cite{KorthalsAltes:2020ryu}.

There is one important difference between the gluon self-energy computed in the presence of
external sources of Polyakov loops, Eq.~(\ref{self_energy_source}),
and those here, Eq.~(\ref{massive_self_energy}).
External sources of Polyakov loops carry zero spatial momentum, and so 
$\Pi_\epsilon^{00}$, while a function of $p_0^{12}$, is independent of the spatial momentum, $p$.

In contrast, dynamical fields carry nonzero spatial momentum, and so produce 
holonomous hard thermal loops, $\delta\Pi^{\mu \nu}(P)$ 
and $\delta\widetilde{\Pi}^{\mu \nu}(P)$.  When $Q$ and $p$ are soft,
up to trivial factors of $\hat{p}^i$ the dimensionless function $\delta\Pi^{\mu \nu}(P)$
is a function of $p/Q$; $\delta\widetilde{\Pi}^{\mu \nu}(P)$ is $1/Q$ times a function of $p/Q$.
We compute these functions shortly and show that all of these functions are nontrivial, Eqs.
(\ref{pi_lg}) - (\ref{tilde_pi_tr}).  

Because the self-energy in Eq.~(\ref{self_energy_source}) is independent of the spatial momentum, though,
it is just $1/p_0^{a b}$ times a function of the $Q$'s.  In the static limit, $p_0 = 0$, it reduces to a function
of the $Q$'s, independent of $p$.  Further, to leading order in $g^2$ when $Q$ is soft
we can just set $Q = 0$, since any $Q \sim g T$ represents a contribution to higher loop order.
Explicit computation demonstrates that when $Q = 0$, 
$\Pi_\epsilon^{00; ab,cd}(Q^a,0)$ reduces to $g^2 T^2$ times a constant, which 
cancels exactly against the usual Debye mass squared
\cite{KorthalsAltes:2019yih, KorthalsAltes:2020ryu}.
This cancellation follows because the total gluon self-energy is transverse:
at zero spatial momentum, $\Pi_{\rm total, \epsilon}^{0 i} = 0$, and so
the first condition in Eq.~(\ref{constraint_transverse}) implies that
$\Pi_{\rm total , \epsilon}^{0 0}(Q^{a b},0) = 0$
when $Q^{a b} \neq 0$.  This contrasts with $p_0 = p = Q =0$,
where $\Pi^{00} \neq 0$, equal to the Debye mass squared, is consistent with transversity when $p_0 = 0$.
This cancellation for soft $Q^a$ is the origin of the conundrum for the free energy $\sim g^3$ in the
presence of an external source: there is no $\sim g^3$ term from off-diagonal gluons,
where $Q^{a b} \neq 0$, since the source terms completely cancel the usual contributions to the Debye mass.

In this way, the addition of new dynamical fields {\it may} solve this conundrum, since then
the gluon self-energy is a function of the spatial momentum.  As we show, however, the condition that the
$\sim g^3$ term in the free energy is smoothly behaved as $Q \rightarrow 0$ is nevertheless a nontrivial
constraint on the auxiliary fields.

\section{Two dimensional ghosts}
\label{sec:two_dim}

\subsection{Embedding two into four dimensions}

In the Introduction, Sec.~(\ref{sec:introduction}),
we argued that inducing nontrivial holonomy through the second Bernoulli polynomial is
very natural.
Besides using massive fields as in the previous section, there is another
way to generate a $B_2(q)$.
Consider a fermionic field in the adjoint representation of a gauge
group: their contribution to the free energy in $d$ spacetime dimensions, at nonzero holonomy and temperature,
involves the polylogarithm of order $d$  \cite{Nishimura:2017crr}.
For even $d$ this polylogarithm reduces to a constant times the Bernoulli polynomial, $B_d(q)$.

Thus one way to generating $B_2$ is to embed a two dimensional field isotropically in four dimensions.  To generate
a free energy which is proportional to $B_2(q)$ 
at nonzero holonomy and temperature, we cannot change the time direction.  That implies we have
to single out one spatial direction.   Thus we
introduce a unit vector $\hat{n}$ in the three spatial directions, and so as not to disturb rotational
symmetry, integrate over all directions of $\hat{n}$.
The longitudinal and transverse coordinates with respect to $\hat{n}$ are
\begin{equation}
  x^i = (\hat{x}, x_\perp) \;\;\; , \;\;\;
  \hat{x} = x \cdot \hat{n} \;\;\; , \;\;\; x_\perp \cdot \hat{n} = 0 \; .
\end{equation}
We then introduce gauge covariant derivatives for the fermionic field, $\phi$:
\begin{equation}
  {\cal S}_{\rm 2D} = \; \int^{1/T}_0 d\tau \int \frac{d\Omega_{\hat{n}} }{4 \pi}
  \int^\infty_{-\infty} d\hat{x} \int_{|x_\perp^2| > 1/C} d^2x_\perp
  \; {\rm tr} \left( (\hat{D} \bar{\phi})(\hat{D} \phi) + (D_\perp \bar{\phi})(D_\perp \phi) \right) \; .
\end{equation}
As for the massive field, $\phi$ and $\bar{\phi}$ are necessarily in the adjoint representation.  
We choose periodic boundary conditions for $\phi$ and
$\bar{\phi}$ in the imaginary time direction, so
that this is necessarily a ghost field. While peculiar, perhaps it isn't so objectionable for an effective theory.

That these fields are ghosts is necessary so that the two-dimensional fields decrease the pressure.
In contrast, the massive fields in the previous section are physical, but by taking a derivative with
respect to the mass squared, Eq.~(\ref{mass_action}), the net pressure.  This can be understood as follows.
The pressure for a massless gas is, of course, positive, $\sim + T^4$.  For a gas whose mass is much less than the temperature,
there is a correction to the pressure, $\sim - m^2 T^2$; this coefficient is negative, as a mass decreases the
pressure.  By taking the derivative with respect to the mass squared, one picks out this correction, which is
then negative.  

For the transverse directions, we only integrate over a sphere about the origin, where
$x_\perp^2 > 1/C$.  
As usual, the imaginary time $\tau$ runs from $0 \rightarrow 1/T$, and the longitudinal direction
$\hat{x}$ from $-\infty$ to $+ \infty$.  That is, the ghost field is two dimensional at short distances,
and four dimensional at large distances.

While we differentiate between the covariant derivatives in the longitudinal and transverse directions,
in and of itself this does not conflict with gauge invariance.  Introducing the cutoff scale $C$
certainly does, as we discuss below.  Thus this model can only be considered as an illustration of a more complete,
and self consistent, theory.  We shall be careful, however, to compute only in limits which manifestly
respect gauge invariance.

There is a natural physical motivation for the behavior of these fields.  At high temperature, the relevant
degrees of freedom are quarks and gluons; at low temperature, there are only confined confined states.  As the temperature is raised
above the critical temperature, for a range of temperature confined states should persist in the deconfined
(or chirally symmetric) phase.  We suggest that this remains
true in a pure gauge theory, even if the deconfining phase transition is of first order.

In a pure gauge theory, the confined states are glueballs, which can be modeled by an effective theory of
strings.  In the absence of dynamical quarks, the strings must be closed, and sweep out two dimensional
surfaces in spacetime.  

How then could closed strings persist in the deconfined phase?  Since it is deconfined,
the breaking of a flux sheet only costs a finite amount of energy.  Thus closed strings form
a two dimensional sheet over short distances, but over distances related to the scale of confinement,
break, and then are fully four-dimensional fields.  

It is easy to see how this action generates the second Bernoulli polynomial.  To one-loop
order, the potential is proportional to
\begin{eqnarray}
  {\cal V}_{\rm ghost} &=& (-) \; {\rm Tr} \; \log \; \left( - \hat{D}^2 - D_\perp^2 \right)\nonumber \\
  &=& (-) \; T \, \int_{-\infty}^\infty \frac{d\hat{k}}{2 \pi} \int_{|k_\perp|<\sqrt{C}} \frac{d^2 k_\perp}{(2 \pi)^2}
      \; \sum_{a,b=1}^N \mathcal{P}^{a b}_{a b}
      \left( \log \left(1 - {\rm e}^{- \sqrt{\hat{k}^2 + k_\perp^2}/T - 2 \pi i q^{ab}} \right)
    + {\rm c.c.}\right) \; .\nonumber\\
\end{eqnarray}
The minus sign comes from the fermionic integral for the ghosts.
The integral over $\hat{k}$ is dominated by momenta $\sim T$; that over $k_\perp$, by those $\sim \sqrt{C}$.
We assume for simplicity that $T \gg \sqrt{C}$, so that we can neglect $k_\perp$ in the energy.  The integral
over $k_\perp$ just gives an overall factor of $C$, with
\begin{equation}
  {\cal V}_{\rm ghost} =
   \; \frac{C\, T^2 }{4 \pi^2} \sum_{a,b=1}^N
  \sum_{n=1}^\infty \frac{1}{n^2} \mathcal{P}^{a b}_{a b} \left( {\rm e}^{- 2 \pi i n q^{ab}} + {\rm c.c.}\right) \; = \; 
  \frac{C \, T^2}{2} \sum_{a,b=1}^N \mathcal{P}^{a b}_{a b}\, B_2(|q^{ab}|_{\mathrm{mod}\; 1}) \; .
\end{equation}
The cutoff was chosen to agree with the result from a massive ghost, Eq.~(\ref{eff_lag_massive}).  

At the outset, we comment that it is possible that two dimensional massless fields give
rise to infrared divergences at nonzero temperature.  Consider the tadpole diagram in
the presence of nonzero $A_0 \sim Q$ in two spacetime dimensions:
\begin{equation}
   \int\frac{d^2K}{(2\pi)^2} \; \frac{1}{(K+Q)^2}
  \; \sim \; \int \; \frac{dk}{k}
  \left( 1 + \frac{1}{{\rm e}^{(k+ i Q)/T} - 1} + \frac{1}{{\rm e}^{(k- i Q)/T} - 1} \right) \; .
  \label{tadpole_diagram}
\end{equation}
This has a logarithmic divergence at zero temperature.  At nonzero temperature, one would
expect this to turn into a linear power divergence.  However, in a holonomous plasma
there is no infrared divergence due to the peculiar identity of Eq.~(\ref{identity_sum}).

Nevertheless, this tadpole diagram does not contribute to the holonomous hard thermal loop (HHTL)
in the gluon self-energy.
In four dimensions, it appears that the contributions to Hard Thermal Loops are of two
types: tadpole diagrams, analogous to Eq.~(\ref{tadpole_diagram}), and those which
contribute to Landau damping in Minkowski spacetime, such as $\mathcal{I}_2$ and
$\mathcal{I}_3$ in Eqs.~(\ref{landau2}) and (\ref{landau3}).
At one-loop order, the diagram with two three gluon vertices produces both terms;
the diagram with one four gluon vertex, only a tadpole diagram.
The tadpole diagrams contribute between these two diagrams, leaving only the diagrams
from Landau damping; this is the reason for the cancellation of the terms $\sim \delta^{i j}$
between Eqs.~(\ref{htl_general}) and (\ref{I14_HTL}) in Eq.~(\ref{final_massive}).
Of course there is a logarithmic infrared divergence at zero temperature from two dimensional
fields, but as an effective theory at nonzero temperature, we ignore this.

The computation of two-dimensional ghosts to the HHTL to the gluon self-energy follows
immediately from previous computations.  Because of our choice of the constant $C$,
the result for ${\cal K}^{ab,cd}$ in Eq.~(\ref{K_result}) and $(m^2_\text{gl})^{ab,cd}$
in Eq.~(\ref{m2_result}) are identical.  Since the two dimensional field is massless, though,
there is no additional contribution to the Debye mass, $(\widetilde{m}^3_\text{gl})^{ab,cd}$
in Eq.~(\ref{tildem2_result}).

\section{Holonomous hard thermal loops and the free energy}
\label{sec:hhtl}

\subsection{Computing HHTL's}
\label{compute_hhtl:sec}

We begin by computing the holonomous hard thermal loops (HHTL's) of Eqs.~(\ref{landauQ_T})
and (\ref{tilde_Pi}).  While our formula's can be used to compute after analytic continuation to
Minkowski energies, we apply our results to the computation of the free energy to $\sim g^3$.
We then consider zero energy in the Euclidean theory, $p_0 = 0$, for soft holonomy, where
all $Q^a$ are soft, $\sim g T$.  For simplicity, we compute as a function of a single $Q$,
where the external momentum is
\begin{equation}
  P^{\mu} = (Q,\vec{p}) \;\; , \;\; P \cdot \widehat{K} = i Q + \vec{p}\cdot \widehat{\vec{k}} = i Q + p \cos\theta \; ,
  \label{p_def}
\end{equation}
the generalization to arbitrary $Q^a$ immediate.

We need to compute two self-energies,
$\delta \Pi^{\mu \nu}$ and $\delta \widetilde{\Pi}^{\mu \nu}$.
From Eqs.~(\ref{pi_transverse}) and (\ref{trans_tilde_pi}),
these are both transverse in a single momentum.  
This allows us to decompose each function into two scalar functions.
For $\delta \Pi^{\mu \nu}$, the longitudinal, $\delta \Pi_\text{lg}$, and
transverse, $\delta \Pi_\text{tr}$, functions are defined as
\begin{eqnarray}
  \delta \Pi^{00}(P) &=& \delta \Pi_\text{lg}(P) \; \; , \;\;
\delta \Pi^{0i}(P) = - \, \hat{p}^i \; \fraqp \; \delta \Pi_\text{lg}(P) \;\; , \;\; \nonumber \\
\delta \Pi^{i j}(P) &=& \left( \delta^{i j} - \hat{p}^i \hat{p}^j \right) \delta \Pi_\text{tr}(P)
+ \hat{p}^i \hat{p}^j \, \frac{Q^2}{p^2} \, \delta \Pi_\text{lg}(P) \; .
\end{eqnarray}
Similarly, from $\delta \widetilde{\Pi}^{\mu \nu}(P)$ we define the longitudinal and transverse
self-energies, $\delta \widetilde{\Pi}_\text{lg}(P)$ and $\delta \widetilde{\Pi}_\text{tr}(P)$.

To compute these functions, we need the angular integrals:
\begin{eqnarray}
  \int \frac{d \Omega}{4 \pi} \; \frac{1}{P \cdot K} &=& - \; \frac{i}{p} \arctan\left( \frapq\right) \; ,
    \label{ang1} \\
  \int \frac{d \Omega}{4 \pi} \; \frac{\cos^2\theta}{P \cdot K}
  &=& - \; \frac{i \, Q}{p^2} \; \left( 1 - \fraqp \arctan\left(\frapq\right) \right) \; ,
\label{ang2} \\
  \int \frac{d \Omega}{4 \pi} \; \frac{1}{(P \cdot K)^2}
  &=& - \; \frac{1}{p^2 + Q^2} \; ,
\label{ang3} \\
  \int \frac{d \Omega}{4 \pi} \; \frac{\cos^2 \theta}{(P \cdot K)^2} &=&
\frac{1}{p^2} \left( 1 + \frac{Q^2}{p^2+Q^2} - 2 \, \fraqp \arctan\left( \frapq \right) \right) \; .
                                                                         \label{ang4} \\
  \nonumber
\end{eqnarray}

Using these integrals, we find
\begin{eqnarray}
  \delta \Pi_\text{lg}(P) &=& -1 + \fraqp \arctan\left(\frapq \right) \; ,
                         \label{pi_lg} \\
  \delta\Pi_\text{tr}(P) &=& \frac{Q}{2p} \left( \fraqp
                        - \left( 1 + \frasqqp \right) \arctan\left(\frapq \right) \right) \; ,
                        \label{pi_tr} \\
  \delta \widetilde{\Pi}_\text{lg}(P) &=& -\frac{1}{Q} \left(
\frac{Q^2}{p^2 + Q^2} -                             
                                     \fraqp \; \arctan\left( \frapq \right) \right) \; ,
                         \label{tilde_pi_lg}  \\
  \delta\widetilde{\Pi}_\text{tr}(P) &=&  \frac{1}{2p} \left( \fraqp
                        + \left( 1 - \frasqqp \right) \arctan\left(\frapq \right) \right) \; .
                        \label{tilde_pi_tr} \\
  \nonumber
\end{eqnarray}

For the known hard thermal loops, $\delta \Pi_\text{lg}$ and $\delta \Pi_\text{tr}$, of course we could have
read off the above simply by taking the known results for Minkowski energy, $p_0 = i \omega$
\cite{Bellac:2011kqa}, and analytically continuing back to Euclidean momenta, taking $\omega = - i Q$.
Doing so, the function $\arctan(p/Q)$ above is related to
$\log((\omega - p)/(\omega + p))$.  Of course this would not yield the new function for an auxiliary
massive field, $\widetilde{\Pi}^{\mu \nu}(P)$.

For the next section, we also need the limits of these self-energies for small and large momenta:
\begin{eqnarray}
  \delta \Pi_\text{lg}(P) & \approx & -\frac{1}{3} \, \frasqpq + \frac{1}{5} \, \fraqupq + \ldots
  \;\; \text{for} \;\; p \ll Q \; ,
                                 \label{small_pi_lg} \\
      & \approx & -1 + \frac{\pi}{2} \, \frac{|Q|}{p} + \ldots \;\;\;\; \quad\text{for} \;\; p \gg Q \; ,
                                 \label{large_pi_lg} \\
  \delta \Pi_\text{tr}(P)
  & \approx & -\frac{1}{3} + \frac{1}{15} \, \frasqpq + \ldots
 \quad \;\,\, \text{for} \;\; p \ll Q \; ,
                                 \label{small_pi_tr} \\
                     & \approx & - \frac{\pi}{4} \, \frac{|Q|}{p} + \frasqqp + \ldots \,\quad \text{for} \;\; p \gg Q \; .
                                 \label{large_pi_tr} \\
  \nonumber
\end{eqnarray}
For the new HHTL,
\begin{eqnarray}
  \delta \widetilde{\Pi}_\text{lg}(P) & \approx &
        \frac{1}{Q} \left( \frac{2}{3} \, \frasqpq - \frac{4}{5} \, \fraqupq + \ldots \right)
  \;\;\qquad \text{for} \;\; p \ll Q \; ,          \label{small_tildepi_lg} \\
                                 & \approx &  \frac{1}{p} \left( {\rm sign}(Q) \frac{\pi}{2} - \fraqp
     + \ldots \right) \; \qquad\text{for} \;\; p \gg Q \; ,    \label{large_tildepi_lg} \\
  \delta \widetilde{\Pi}_\text{tr}(P) & \approx & \frac{2}{3 Q} \left( 1 - \frac{2}{5} \, \frasqpq + \ldots \right)
  \;\;\, \quad\qquad \text{for}\;\; p \ll Q \; ,
                                 \label{small_tildepi_tr} \\
                                 & \approx & {\rm sign}(Q) \frac{\pi}{4p} \,\left( 1 - \frasqqp + \ldots \right)
                                             \; \quad\text{for} \;\; p \gg Q \; .
                                 \label{large_tildepi_tr} \\
  \nonumber
\end{eqnarray}

For soft $Q \sim g T$, we can compute the coefficients at $Q = 0$, which gives:
\begin{equation}
  (m^2_\text{gl})^{ab,cd}(0) = m_D^2 \, {\cal P}^{ab,cd} \;\; , \;\; m_D^2 = g^2 N \left(\frac{T^2}{3}  + \frac{C}{4 \pi^2} \right)
  \;\; ,
  \label{debye_mass}
  \end{equation}
where $m_D$ is the usual Debye mass for static electric fields, generalized to $C \neq 0$.  The mass scale associated
with the auxiliary massive field is
  \begin{equation}
    (\widetilde{m}^3_\text{gl})^{ab,cd}(0,0) = - \frac{ g^2 N \, C  \, T}{32 \pi} \; .
    \label{tilde_mass}
  \end{equation}
There is no simple understanding for $\widetilde{m}_{\rm gl}^3$, which has dimensions of mass cubed
(remember $C$ has dimensions of mass squared).

\section{Free energy to cubic order}
\label{sec:free_energy}

\subsection{General Expressions}

At nonzero holonomy, the computation of the free energy to $\sim 1$ \cite{Gross:1980br,Weiss:1980rj}
and $\sim g^2$
\cite{Belyaev:1991gh,Belyaev:1991np,Smilga:1993vb,Kogan:1993bz,bhattacharya_interface_1991,bhattacharya_zn_1992,Sawayanagi:1992as,Sawayanagi:1994nz,Skalozub:1992un,Skalozub:1992pb,Skalozub:1994br,Skalozub:1994pa,Borisenko:1994jn,Skalozub:2004ab,Bordag:2018aii,Borisenko:2020dej,Skalozub:2020idf,KorthalsAltes:1993ca,Giovannangeli:2002uv,Giovannangeli:2004sg,Dumitru:2013xna,Guo:2014zra,Guo:2018scp}
is straightforward.  This is because the dominant momenta are on the order of the temperature, and so these
are well behaved for any value of the holonomy.

At zero holonomy, there are infrared divergences from the static modes, with $p_0 = 0$, which
first appear at $\sim g^4$.  These infrared
divergences are cut off by a nonzero value for the Debye mass, and after resummation,
generate a term which is $\sim g^3$, as computed first by Kapusta \cite{Kapusta:1979fh}.
When the holonomy is nonzero but soft, $Q \sim g T$, the holonomy is as large as the Debye mass, and a
nontrivial function results.  The purpose of this section is to see under which conditions the
terms $\sim g^3$ are well behaved as the holonomy vanishes.

Our purpose is to see if the cubic term in the free energy behaves smoothly as $Q \rightarrow 0$, and so
we consider only the contribution from a single mode with holonomy $Q$.  This is the contribution of
off-diagonal gluons to the free energy for two colors.  The generalization to higher number of colors is
immediate, and so we suppress the color indices.  We comment that the result will be proportional
to the Debye mass cubed, and so survives as the number of colors $N \rightarrow \infty$.
We also concentrate on the contribution only from the terms involving the Debye mass,
$\sim \delta \Pi^{\mu \nu}$, and comment later on that from the new piece from the auxiliary
massive mode, $\sim \delta \widetilde{\Pi}^{\mu \nu}$.

The free energy to $\sim g^3$ is gauge invariant because of the
HHTL is transverse in the external momentum, Eq.~(\ref{pi_transverse}) and (\ref{trans_tilde_pi}).
We choose to work in Feynman gauge.

The contribution from transverse gluons is
\begin{eqnarray}
  {\cal F}_3^{\rm tr}(Q/m_D) &=& T \sum_{n=-\infty}^{+\infty} \int \frac{d^3p}{(2 \pi)^3} \;
                   \left(
    \log \left( (p_0 + Q)^2 + p^2 - m_D^2 \; \delta \Pi_{\rm tr}(p_0 + Q,p) \right) \right. \nonumber \\
    &-& \left. \log \left( (p_0 + Q)^2 + p^2 \right) + \frac{m_D^2 }{(p_0 + Q)^2 + p^2 }\; \delta\Pi_{\rm tr}(p_0+Q,p) 
  \right)     \; , \nonumber \\
  \label{trans_gcubed}
\end{eqnarray}
where $m_D^2$ is that of Eq.~(\ref{debye_mass}).
The overall coefficient
is due to $2$ from transverse modes times $1/2$ for a bosonic field.  In covariant gauges, 
the $2$ comes from $4$ gluon modes minus $2$ ghosts.

To obtain a term $\sim g^3$, it is necessary to first subtract the terms which arise at lower order.
There are two such terms.  The first arises at one loop order, $\sim 1$, which is the second term on
the right hand side of Eq.~(\ref{trans_gcubed}),
$\sim {\rm tr} \log( (p_0 + Q)^2 + p^2)$.  Next is the term at two loop order,
$\sim m_D^2 \delta \Pi_{\rm tr} \sim g^2$, which is the last term in Eq.~(\ref{trans_gcubed}).

The term $\sim g^3$ arises from the static mode of the inverse propagator, taking
$p_0 = 0$ in Eq.~(\ref{trans_gcubed}), and thus is
\begin{equation}
  {\cal F}_3^{\rm tr}(Q/m_D) = T \int \frac{d^3p}{(2 \pi)^3} \; \left(
    \log \left( 1 - \frac{m_D^2}{Q^2 + p^2} \; \delta \Pi_{\rm tr}(Q,p) \right)
    + \frac{m_D^2 }{Q^2 + p^2 } \; \delta\Pi_{\rm tr}(Q,p) 
  \right) 
    \; .
  \label{trans_gcubed_simple}
\end{equation}

For the longitudinal propagator the time and spatial components of the
propagators mix.  For simplicity, assume that $p^i$ is along the $z$ direction.  Anticipating our results,
we work in the static limit, $p_0 = 0$.  The inverse propagator is a two by two matrix,
\begin{equation}
  \Delta^{-1} =   \left(
\begin{array}{cc}
  Q^2 + p^2 - m_D^2 \, \delta \Pi_{\rm lg} &  m_D^2 \, \delta \Pi_{\rm lg} \; Q /p \\
   m_D^2 \, \delta \Pi_{\rm lg}\; Q/p & \;\; Q^2+p^2 - m_D^2 \, \delta \Pi_{\rm lg}\; Q^2/p^2  \; \end{array}
\right) \; .
\label{inverse_prop}
\end{equation}
The determinant of this matrix is
\begin{equation}
  \det \Delta^{-1} = (Q^2 + p^2)^2 \left( 1 - \frac{m_D^2}{p^2} \; \delta \Pi_{\rm lg} \right) \; .
\end{equation}
Consequently, the contribution of the longitudinal modes to the free energy at cubic order is
\begin{equation}
  {\cal F}_3^{\rm lg}(Q/m_D) = T  \int \frac{d^3p}{(2 \pi)^3} \left(
    \log \left( 1- \frac{m_D^2}{p^2} \; \delta \Pi_{\rm lg}(Q,p) \right)
    + \frac{m_D^2 }{p^2 }\; \delta\Pi_{\rm lg}(Q,p)
  \right) 
    \; .
  \label{trans_gcubed_long}
\end{equation}

The functions in Eqs.~(\ref{trans_gcubed_simple}) and (\ref{trans_gcubed_long}) are
$\sim T m_D^3$ times a dimensionless function of $Q/m_D$.  The integrals are
well defined and convergent in both the ultraviolet and infrared limits, and so can be
determined numerically.  First we determine their values in the limit of small and large holonomy.

\subsection{Limits}

For zero holonomy, using the integral
\begin{equation}
  \int^{\infty}_0 dp \; \left( p^2 \; \log \left(1 + \frac{1}{p^2} \right)  - 1 \right)
 = - \, \frac{\pi}{3} \; ,
\label{zero_holonomy}
\end{equation}
we find the standard result \cite{Kapusta:1979fh},
\begin{equation}
  {\cal F}_3^{\rm lg}(0) = - \frac{1}{6 \pi} \, m_D^3 \, T \;\; , \;\;
  {\cal F}_3^{\rm tr}(0) = 0 \; .
  \label{free_zero}
\end{equation}

We begin with small holonomy, where $Q \ll m_D$.  The dominant momenta are then
\begin{equation}
  Q \ll p \ll m_D \; .
\end{equation}
That is, and somewhat counterintuitively, in order to obtain the behavior for small $Q$, we need the behavior
of the self-energies in the limit of large spatial momentum, $p \gg Q$.
Thus the usual result for the Debye mass squared at zero holonomy is given by $\delta \Pi_{\rm lg} \sim -1$ at large $p$,
Eq.~(\ref{large_pi_lg}), which gives Eq.~(\ref{free_zero}).

To compute corrections to this result, we need to expand $\delta \Pi_{\rm lg}$
in Eq.~(\ref{large_pi_lg}) to linear order in $Q$.   Substituting this into Eq.~(\ref{trans_gcubed_long}),
to $\sim Q$
\begin{eqnarray}
  {\cal F}_3^{\rm lg}(Q/m_D) - {\cal F}_3^{\rm lg}(0) &\approx&
 \frac{m_D^2\, T}{2 \pi^2} \int dp \; p^2 \; \left(\frac{\pi \, Q }{2\, p}\right)
              \left( \frac{1}{p^2} - \frac{1}{p^2 + m_D^2} \right)
 \nonumber \\
                                                      &\approx& \frac{m_D^2\, T}{4 \pi} \; Q \; \left(
        \log\left(\frac{m_D}{Q}\right) + O(1) \right) + O(Q^2) \;\;\quad \text{for} \;\; Q \ll m_D \; . \nonumber\\
  \label{long_small_holon}
\end{eqnarray}
This is valid for $Q \ll g T$, and so overall is $\sim m_D^2 Q T\ll g^3 T^4$ in magnitude.

For the transverse modes, the self-energy at high momentum is given by Eq.~(\ref{large_pi_tr}).  Because this
vanishes at high momentum, the terms of $\sim Q$ are given by expanding the self-energy to linear order, and cancel
identically.  There are contributions from the transverse modes to $\sim Q^2$.

For large holonomy, $Q \gg m_D$, 
consider first the contribution of the transverse modes.
Because the transverse self-energy $\delta \Pi_{\rm tr}$ in Eq.~(\ref{trans_gcubed})
is accompanied by a factor of $1/(Q^2 + p^2)$,
for large $Q$ we can expand to quadratic order in 
$\delta \Pi_{\rm tr}/(Q^2 + p^2)$.  This factor of $1/(Q^2 + p^2)$ ensures that the dominant
momenta for the transverse free energy are $m_D \ll p \ll Q$, and so we can expand the transverse
self-energy for small momenta, Eq.~(\ref{small_pi_tr}).  We only
need the leading term, $\delta \Pi_{\rm tr}(p/Q) \approx -1/3$, which gives 
\begin{equation}
  {\cal F}_3^{\rm tr}(Q/m_D) \approx \frac{T}{2 \pi^2} \int dp \; p^2 \left( - \frac{1}{2} \right)
  \left(\frac{1}{ Q^2 + p^2} \frac{(-)m_D^2}{3}  \right)^2 = - \frac{T}{144 \pi} \, \frac{m_D^4}{Q} + \ldots \;\quad \text{for}
  \; \; Q \gg m_D \; .
  \label{trans_large_hol}
\end{equation}

We comment that the limit of small $p \ll Q$ corresponds to the limit of nonzero frequency and zero spatial
momentum at zero holonomy.  This explains why the leading term in Eq.~(\ref{small_pi_tr}) is $1/3$ the value
in Eq.~(\ref{large_pi_lg}), as $\Pi^{\mu \mu}(0,0)$ is the same for the limits of $p_0 = 0$ and $p \rightarrow 0$
and $p=0$, $p_0 = i \omega$, $\omega \rightarrow 0$.

For the longitudinal modes the analysis is slightly more subtle.  In this case, the longitudinal
self-energy is multiplied by $1/p^2$, not $1/(Q^2 + p^2)$, as for the transverse case.
For large $Q$, we can still expand to quadratic order in $\delta \Pi_{\rm lg}/p^2$:
\begin{equation}
  {\cal F}_3^{\rm lg}(Q/m_D) \approx \frac{T}{2 \pi^2} \int dp \; p^2 \left( - \frac{1}{2} \right)
  \left( \frac{m_D^2}{p^2} \; \delta \Pi_{\rm lg}\left(Q,p \right) \right)^2  + \ldots \;  \quad\text{for} \; \; Q \gg m_D \; .
\end{equation}
In this expression the Debye mass only enters through an overall factor of $\sim m_D^4$,
leaving an integral,
\begin{equation}
  {\cal F}_3^{\rm lg}(Q/m_D) \approx - \frac{m_D^4 \, T}{4 \pi^2 } \int^\infty_0 dp \; \frac{1}{p^2}
  \left(\delta\Pi_{\rm lg}\left(Q,p\right) \right)^2
  =  \frac{T}{12 \pi } \; \frac{m_D^4}{Q} \; \left( \log(2) - 1 \right) \; \quad \text{for} \;\;
  Q \gg m_D \; .
  \label{long_large_hol}
\end{equation}
The dominant momenta  in the integral are $p \sim Q$ and so we need the complete expression for
$\delta \Pi_{\rm lg}$ in Eq.~(\ref{pi_tr}).  
Nevertheless, the coefficient of the term $\sim 1/Q$ is just an integral over the longitudinal self-energy
which can be done exactly.

That the contributions at large holonomy vanish as $\sim 1/Q$ for the both the transverse and longitudinal modes
is hardly surprising.  There are no infrared divergences when $Q$ is large, and so the contributions in both
Eqs.~(\ref{trans_large_hol}) and (\ref{long_large_hol}) are simple to determine, given directly by expanding
the expression for the free energy to quadratic order in the $\delta \Pi$'s.  As such, they are just part of the
usual, perturbative contribution to the free energy, with terms $\sim m_D^4$ just part
of those $\sim g^4$, at three loop order.

Having derived the results for two dimensional ghosts, it is immediate to include the
results for an auxiliary massive field.  Consider the form of the longitudinal self-energy at large momenta,
which we argued above is relevant for small $Q$, Eq.~(\ref{long_small_holon}).
For $p \gg Q$, the total longitudinal self-energy for a massive auxiliary field is
\begin{equation}
  \Pi_{\rm total ; \rm lg} \approx m_D^2 - \frac{\pi}{2 p} (m_D^2 Q + \widetilde{m}_D^3 ) + \ldots \; .
\end{equation}
As $p \rightarrow \infty$, this equals the Debye mass squared, as expected.  However, consider the behavior
of the leading correction, when both $p$ and $Q$ are soft, $\sim g T$.  Then the first term is
$m_D^2 Q/p \sim g^2 T^2$, which is of the same order as the leading term.  However, the second term
is $\widetilde{m}_D^3/p \sim g^2 T^3/p \sim g T^2$ when the spatial momentum is soft,
$p \sim g T$.  This violates the usual
power counting of hard thermal loops, where the self-energy is as large as the terms at tree level.
For example, the correction to the free energy in Eq.~(\ref{long_small_holon}) becomes
\begin{equation}
    {\cal F}_3^{\rm lg}(Q) - {\cal F}_3^{\rm lg}(0) \approx
    \frac{T}{4 \pi} \log\left(\frac{m_D}{Q}\right)
    \left( m_D^2 Q + \widetilde{m}^3_D \right) + \ldots \quad \text{for} \;\; Q \ll m_D \; .
    \label{modified_long_small_holon}
  \end{equation}
  Thus the correction to the free energy from the longitudinal mode does not vanish smoothly as $Q \rightarrow 0$
  when $\widetilde{m}_D \neq 0$.

Clearly we have only computed part of the free energy to $\sim g^3$.  In particular, any boson field which is
originally massless will acquire a thermal mass squared $\sim g^2 T^2$; the associated mode
with zero energy, $p_0 = 0$, then contributes to the free energy at $\sim g^3$.  Why, then, do we concentrate
only upon the gluon contribution?  The example of massless quarks at nonzero holonomy shows that
their contribution has no anomalous terms as arise for gluons \cite{Hidaka:2009hs}, and surely the same
is true for additional scalar fields.   Uniquely, the only place where the one point function enters is for the
gluon self-energy, as a measure of the total color current.  Consequently, we expect that it is only for
gluon fields that there is a problem with the self-energy as $Q \rightarrow 0$.

This problem is not special to the hard thermal loop limit for Euclidean momenta.
If one analytically continues the self-energies to Minkowski momenta, $p_0 \rightarrow i \omega$ for
soft $\omega \sim g T$, the new HHTL $\delta \widetilde{\Pi}$ has terms $\sim g T^2$,
instead of the expected $\sim g^2 T^2$.  
  
It is important to acknowledge
that our effective theories are manifestly incomplete.  For either the theory with auxiliary
massive fields, or embedded two dimensional fields, the self-energies of their additional fields are
gauge variant.  This is easiest to see for the latter.  The thermal mass of a scalar
field is due to two contributions, from a tadpole diagram, involving a gluon loop, and a second diagram,
with a discontinuity from a virtual scalar-gluon intermediate state.  In four dimensions, each diagram
has a piece which depends upon the gauge fixing parameter, and they cancel between the two.  For the
effective two dimensional theory, however, the tadpole diagram is unchanged, as it only involves a gluon
loop, while for the second diagram the momentum for the scalar is modified, and so the cancellation fails.
Thus there are additional contributions to ensure that the total free energy is gauge invariant.  For
the reasons discussed above, however, we do not expect this to change the behavior as the holonomy
$Q \rightarrow 0$.

\section{Conclusions}

As discussed in the Introduction, to avoid an unwanted first order phase transition in the deconfined
phase, the quark gluon plasma must always be holonomous at any finite temperature
\cite{Dumitru:2012fw,KorthalsAltes:2020ryu}.  The simplest way to do this is if
effective fields generate the second Bernoulli polynomial, as that is linear in the holonomy for small
values, Eq.~(\ref{potential_b2}).

We investigated two ways of generating such a term: through auxiliary massive fields, and the isotropic
embedding of two dimensional fields into four dimensions.  In each case, the computation of the gluon
self-energy is well defined.  We computed the behavior when the spatial momenta and the holonomy
are both soft, $\sim g T$.  The self-energy for auxiliary massive fields acquires a new term whose
behavior is $\sim g^2 T^3/p$, where $p \sim g T$ is a soft momentum.  This term is not $\sim g^2 T^2$, as
expected for a consistent effective theory, but is $\sim g T^2$.

This term does not arise for the two dimensional ghosts.  Indeed, computing with these two dimensional
fields is extremely simple: one uses the usual holonomous hard thermal loops (HHTL), but with a propagator
whose Debye mass squared includes the effect of the ghosts, Eq.~(\ref{debye_mass}).  It is very direct
to compute with this effective propagator.  Previously, we computed the shear viscosity in Refs.
\cite{Hidaka:2008dr} and \cite{Hidaka:2009ma}.  This used a HHTL propagator, where the term from
the equations of motion, $\sim {\cal K}^{ab,cd} \delta \Gamma^{\mu \nu}(P^{ab})$, Eqs.
(\ref{landauQ_T}), (\ref{massive_self_energy}), and (\ref{K_result}), was simply dropped by hand.
The present models demonstrate that this is consistent.  Notably, the shear viscosity does decrease
as $T \rightarrow T_c$, due to the decrease in the effective number of degrees of freedom as one
approaches the confined phase \cite{Hidaka:2008dr,Hidaka:2009ma}

Thus improving this result is simply a matter of using the Debye mass squared of Eq.~(\ref{debye_mass}).
In particular, computing the ratio of the bulk to the shear viscosity is 
straightforward.  As a ratio, this should be less sensitive to the various limitations of our approximations.
This computation will be presented separately \cite{hidaka_new}.

These calculations are clearly of use for phenomenology.  We conclude by noting a point of principle.
We argue that the quark gluon plasma is always holonomous, so that over large distances, the self-energy
of the longitudinal fields are unscreened.  This can be seen from Eq.~(\ref{small_pi_lg}), which
vanishes as $\sim p^2/Q^2$.  Taken at face value, then, it appears as if static electric fields are
not screened over large distances.  Since this distance is
$1/Q \sim T/T_c^2$, this may be a very large distance indeed, and extremely difficult to
measure through numerical simulations on the lattice.  Certainly,
it is necessary to look at the T-odd part of Polyakov loops, such as the imaginary part for three
or more colors, as noted by Arnold and Yaffe \cite{Arnold:1995bh}.  While suggested by the perturbative
analysis, however, we suggest that nonperturbative effects {\it may} generate a finite correlation
for static electric fields at nonzero holonomy, by interacting with the dynamics responsible for the
holonomy in the first place.  This is speculative, but it demonstrates that careful analysis
of correlation lengths even in the static, Euclidean theory may yield insight into both the perturbative
and nonperturbative effects in a holonomous quark gluon plasma.

\acknowledgments
This work was supported by JSPS KAKENHI Grant Numbers~17H06462 and 18H01211.
R.D.P. thanks the U.S. Department of Energy for support
under contract DE-SC0012704.  R.D.P. thanks C. Korthals-Altes, H, Nishimura, and V. Skokov
for discussions.

\bibliography{ghosts,qks}

\end{document}